\documentclass[english,11pt, draftclsnofoot, journal, a4paper, oneside, onecolumn]{IEEEtran}
\usepackage[T1]{fontenc}
\usepackage[latin9]{inputenc}
\usepackage{verbatim}
\usepackage{bm}
\usepackage{amsmath}
\usepackage{amssymb}
\usepackage{graphicx}
\usepackage{esint}

\makeatletter

\newcommand{\lyxmathsym}[1]{\ifmmode\begingroup\def\b@ld{bold}
  \text{\ifx\math@version\b@ld\bfseries\fi#1}\endgroup\else#1\fi}

  \newtheorem{asm}{Assumption}
  \newenvironment{asmQED}{\begin{asm}}{~\hfill\IEEEQEDclosed\end{asm}}
  \newtheorem{defQED}{Definition}
  \newenvironment{lyxDefQED}{\begin{defQED}}{~\hfill\IEEEQEDclosed\end{defQED}}
  \newtheorem{problem}{Problem}
  \newtheorem{lemQED}{Lemma}
  \newenvironment{lyxLemQED}{\begin{lemQED}}{~\hfill\IEEEQEDclosed\end{lemQED}}
  \newtheorem{thmQED}{Theorem}
  \newenvironment{lyxThmQED}{\begin{thmQED}}{~\hfill\IEEEQEDclosed\end{thmQED}}
  \newtheorem{corQED}{Corollary}
  \newenvironment{lyxCorQED}{\begin{corQED}}{~\hfill\IEEEQEDclosed\end{corQED}}
  \newtheorem{remrk}{Remark}

\usepackage{color}
\usepackage{cite}

\author{Junting~Chen,~\IEEEmembership{Student~Member,~IEEE}        and~Vincent~K.~N.~Lau,~\IEEEmembership{Fellow,~IEEE}\\
Dept. of Electronic and Computer Engineering \\The Hong Kong University of Science and Technology\\Clear Water Bay, Kowloon, Hong Kong\\ Email: \{eejtchen, eeknlau\}@ust.hk

\thanks{Copyright (c) 2011 IEEE. Personal use of this material is permitted.         However, permission to use this material for any other purposes must         be obtained from the IEEE by sending a request to pubs-permissions@ieee.org.}%

}

\makeatother

\usepackage{babel}
\begin{document}

\title{Delay Analysis of Max-Weight Queue Algorithm for Time-varying Wireless
Adhoc Networks - Control Theoretical Approach}

\maketitle

\begin{abstract}
\textmd{Max weighted queue (MWQ) control policy is a widely used cross-layer
control policy that achieves queue stability and a reasonable delay
performance. In most of the existing literature, it is assumed that
optimal MWQ policy can be obtained instantaneously at every time slot.
However, this assumption may be unrealistic in time varying wireless
systems, especially when there is no closed-form MWQ solution and
iterative algorithms have to be applied to obtain the optimal solution.
This paper investigates the convergence behavior and the queue delay
performance of the conventional MWQ iterations in which the channel
state information (CSI) and queue state information (QSI) are changing
in a similar timescale as the algorithm iterations. Our results are
established by studying the stochastic stability of an equivalent
virtual stochastic dynamic system (VSDS), and an extended Foster-Lyapunov
criteria is applied for the stability analysis. We derive a closed
form delay bound of the wireless network in terms of the CSI fading
rate and the sensitivity of MWQ policy over CSI and QSI. Based on
the equivalent VSDS, we propose a novel MWQ iterative algorithm with
compensation to improve the tracking performance. We demonstrate that
under some mild conditions, the proposed modified MWQ algorithm converges
to the optimal MWQ control despite the time-varying CSI and QSI. }
\end{abstract}
\maketitle
\IEEEpeerreviewmaketitle

\section{Introduction}

Recently, there has been intense research interest studying cross-layer
resource allocation of wireless adhoc networks for delay-sensitive
applications. While the CSI indicates the \emph{transmission opportunity},
the \emph{queue-state-information} indicates the \emph{urgency} of
the packets in the queues. A good control policy (in delay sense)
should strike a balance between the opportunity (CSI) and the urgency
(QSI) and the design is highly non-trivial \cite{Berry02,Bettesh06,Munish03,Wei06,Georgiadis2006,Neely06-Energy}.
One approach, namely the \emph{Lyapunov Optimization} technique \cite{Georgiadis2006,Neely06-Energy},
allows a potentially simple control policy which adapts to the CSI
and QSI. Specifically, the authors in \cite{Georgiadis2006,Neely06-Energy}
have proven that a \emph{max weighted queue} (MWQ) throughput optimization
solution can maximize the negative Lyapunov drift in the queue dynamics
and it can achieve queue stability%
\footnote{Using the MWQ solution, the system of queues can be stable if the
arrival rates are within the \emph{stability region} \cite{Georgiadis2006}
of the systems.%
} with reasonable delay performance. 

In most of the existing literature, it has been commonly assumed that
the MWQ policy can be solved efficiently at each time slot based on
the current realizations of CSI and QSI. However, this assumption
may not be practical for moderate to large scale networks. Specifically,
the MWQ solution requires solving a queue-weighted optimization problem
\cite{Georgiadis2006,Neely06-Energy} and there is no closed-form
solution in most cases. As a result, iterative algorithms (such as
primal dual iterations) have to be used to obtain the MWQ solution
at each time slot. While there is a lot of standard literature establishing
the convergence of the iterative optimization algorithms, these works
have assumed that the CSI and the QSI remains unchanged during the
algorithm iterations%
\footnote{In other words, it is assumed that the algorithm iteration time scale
is much smaller than the CSI / QSI time scale.%
}. However, for large scale networks, the algorithm iteration may involve
not only the node itself but also over-the-air signaling between nodes.
In this case, the CSI and the QSI may have changed after a few iterations
and the existing convergence results (for static problems) failed
to apply in this case of time-varying CSI and QSI. Furthermore, when
the nodes in the adhoc network have limited power and computational
resources, it may not be cost-effective for the node to iterate many
times locally at each time slot as well. These observations motivate
us to study the design and delay analysis of MWQ solutions in time
varying wireless adhoc networks. 

In this paper, we consider a time-varying wireless adhoc network with
power control driven by the MWQ algorithm. We study how the average
delay performance of the MWQ solution is affected by the time-varying
CSI and QSI. Unlike conventional works, we focus on the case where
the MWQ algorithm iteration evolves in a similar timescale as the
CSI and QSI dynamics. There are various first order technical challenges
that have to be addressed. 
\begin{itemize}
\item \textbf{Nonlinear Stochastic Algorithm Dynamics:} One approach is
to adopt continuous time control theory and model the algorithm dynamics
using deterministic ODE \cite{Holliday04,Paul04,Karthik08,LMS2007,Tracking2009}.
In \cite{Holliday04,Paul04,Karthik08}, the authors have considered
the convergence behavior of the Foschini-Miljanic power control algorithm
under time varying channels using the linear ordinary differential
equation (ODE) approach. The authors in \cite{LMS2007,Tracking2009}
studied the tracking performance of the linear least mean square (LMS)
algorithms under time varying channels. However, all these works have
assumed \emph{linear and deterministic} algorithm dynamics and these
approaches cannot be easily extended to our case where the MWQ algorithm
iteration is \emph{nonlinear and stochastic}. 
\item \textbf{Coupled Queue Dynamics and Algorithm Dynamics: }The evolution
of QSI depends on the control actions of the MWQ algorithm in each
time slot. On the other hand, the evolution of the MWQ algorithm also
depends on the time-varying QSI because the MWQ solution is obtained
by solving a queue-weighted optimization. As a result of this mutual
coupling, the techniques in our previous works \cite{Chen2011}, which
considered algorithm tracking performance where control decisions
were made only based on CSI, cannot be easily extended in this case%
\footnote{In the previous work \cite{Chen2011}, the control actions were made
only based on the CSI. However, in this work, we consider CSI and
QSI adaptive control policies and focus on the impact of \emph{both}
the time-varying CSI and QSI on the convergence of the algorithm.
Here, there is a coupled dependency between the control actions (which
depends on QSI) and the time-varying QSI (which depends on the control
actions). This coupled dependency makes the problem challenging. %
}. 
\item \textbf{Delay Analysis and Compensation with Algorithm Tracking Errors:
}It is also quite challenging to analyze and compensate for the delay
penalty due to the MWQ algorithm tracking errors on the power control
actions. In \cite{Georgiadis2006,Neely06-Energy}, the authors have
derived an average delay bound for MWQ algorithm with i.i.d. CSI based
on the Lyapunov drift analysis. However, this technique cannot be
easily extended to our case when there are tracking errors in the
MWQ control actions and correlations in the CSI evolutions. 
\end{itemize}

In this paper, we adopt a continuous time approach to model the algorithm
dynamics of the MWQ power control iterations. We consider Markovian
source arrivals and CSI evolutions so that the combined CSI, QSI and
algorithm dynamics can be modeled by a \emph{stochastic differential
equation} (SDE). We show that studying the convergence behavior in
the \emph{algorithm domain} is equivalent to studying the \emph{stability}
property of a \emph{virtual stochastic dynamic system} (VSDS). Using
non-linear control theory and \emph{stochastic Foster-Lyapunov} techniques,
we establish a bound on delay performance due to time varying CSI
and random source arrivals. Based on the VSDS dynamics, we propose
a modification to the standard MWQ algorithm to compensate for the
penalty due to the time-variation in the wireless adhoc networks.
This paper provides a theoretical framework for studying the convergence
of iterative algorithms as well as potential compensation techniques.
The convergence analysis of iterative algorithms have widespread applications
in network optimizations \cite{Munish03,Georgiadis2006,Neely06-Energy}
and signal processing \cite{LMS2007,Tracking2009}.

\emph{Notations:} $A^{T}$ ($\mathbf{a}^{T}$) denotes the transpose
of matrix (vector) $A$ ($\mathbf{a}$) and $A^{H}$ denotes the complex
conjugate transpose. $|x|$ denotes the absolute value of $x$ and
$\|\mathbf{x}\|=\max_{i}\{x_{i}\}$ denotes the $L_{\infty}$ norm
of vector $\mathbf{x}$. For a complex variable $z$, $\mbox{Re}\left[z\right]$
denotes its real part and $\overline{z}$ denotes its complex conjugate.

\section{System Model and Virtual Stochastic Dynamic Systems\label{sec:System-Model-and-VSDS}}

In this section, we shall first introduce the system model of the
wireless adhoc network as well as the MWQ algorithm. Next, we shall
introduce the notion of \emph{virtual stochastic dynamic system }and
establish the equivalence between the convergence behavior of the
gradient algorithm and the \emph{virtual stochastic dynamic system.}

\subsection{Network Topology, CSI and QSI Models\label{sub:Network-Topology,-CSI}}

\begin{figure}
\begin{centering}
\includegraphics[scale=0.45]{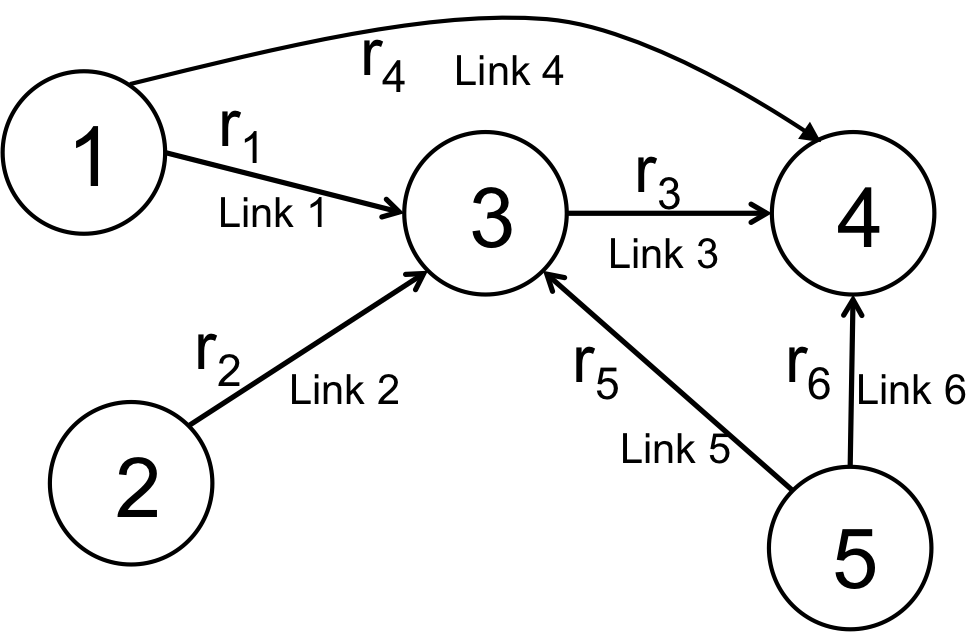}
\par\end{centering}

\caption{\label{fig:Network-topology}Network topology. We consider a wireless
adhoc network with $N$ nodes and $L$ links. We illustrate here $N=5$
and $L=6$ as an example. The $l$-th link transmits the $l$-th data
flow. Transmission flows towards a same destination share the same
frequency band and MUD and SIC are implemented at each receiving node
to handle the inter-flow interference. }
\end{figure}

We consider a wireless adhoc network with $N$ nodes and $L$ links,
where each link corresponds to one transmitting and receiving pair,
as illustrated in Fig. \ref{fig:Network-topology}. Different receiving
nodes occupy different frequency bands, while transmission flows towards
a same destination share the same frequency band. \emph{Multiuser
detection} (MUD) and \emph{successive interference cancellation} (SIC)
are implemented at each receiving node to handle the \emph{inter-flow
interference}. The maximum achievable transmission rate at receiving
node $n$ is a set of rates $\mu_{l}$ that satisfy the following
conditions \cite{Tse2005:fundamental:Wireless}, 
\begin{equation}
\sum_{l\in\mathcal{S}(n)}\mu_{l}<\log\left(1+\sum_{l\in\mathcal{S}(n)}\left|h_{l}\right|^{2}p_{l}\right)\qquad\forall\mathcal{S}(n)\subset\mathcal{L}_{rev}(n)\label{eq:capacity-region}
\end{equation}
where $\mathcal{L}_{rev}(n)$ is a collection of links whose destinations
are at node $n$, $\mathcal{S}(n)$ are any non-empty subsets of $\mathcal{L}_{rev}(n)$,
$h_{l}$ is the channel fading coefficients of link $l$ and $p_{l}$
is the normalized power allocated at link $l$. For example, Fig.
\ref{fig:Network-topology} illustrates an example adhoc network with
$N=5$ nodes and $L=6$ links. $\mathcal{L}_{rev}(3)=\left\{ 1,2,5\right\} ,\mathcal{L}_{rev}(4)=\left\{ 3,4,6\right\} $.
Subscript $l$ represents the link index as well as the flow index.

We have the following assumptions regarding the channel state (CSI)
$h_{l}(t)$. 
\begin{asmQED}
[Temporally Correlated CSI Model]\label{asm:CSI-Model-The} The
support of the channel state process $h_{l}(t)$ is assumed to be
$h_{l}\in\mathcal{H}=\{h\in\mathbb{C}:|h|\geq h_{0}\}$ for some positive
$h_{0}$. Furthermore, $h_{l}(t)$ is a stochastic process described
by the following reflective \emph{stochastic differential equation}
(SDE) in \cite{Skorohod61} 
\begin{equation}
dh_{l}=-\frac{1}{2}a_{l}h_{l}dt+a_{l}^{\frac{1}{2}}dw_{l}+dv_{l},\quad h_{l}\in\mathcal{H}\label{sde:CSI-model}
\end{equation}
where $a_{l}$ determines the temporal correlation of $h_{l}(t)$,
$w_{l}(t)$ is the standard complex Wiener process with unit variance,
and $dv_{l}$ is the Skorohod reflection \cite{Skorohod61} term that
satisfies $dv_{l}(t)\geq0$ and $\int_{0}^{\infty}1\{|h_{l}(t)|>h_{0}\}dv_{l}(t)=0$.
The fading process is independent w.r.t. the link index $l$. 
\end{asmQED}

Note that $h_{l}(t)$ in (\ref{sde:CSI-model}) is a continuous time
version of the auto-regressive (AR) process which has been widely
used to model the dynamics of a correlated wireless fading channel
\cite{Feng2007}. It captures the CSI variation speed that affects
the convergence behavior of the algorithm in a time-varying channel.
It can be shown that the process $h_{l}(t)$ has a stationary distribution. 

%

Incoming data packets randomly arrive at different nodes and are queued
according to their destinations associated with particular transmission
links. Let $q_{l}(t)$ and $N_{l}(t)$ be the current queue backlog
of queue and the number of packets arrived, respectively, at the $l$-th
queue at time $t$. We have the following assumptions regarding the
bursty arrival process $N_{l}(t)$.
\begin{asmQED}
[Bursty Source Model]\label{asm:Bursty-Source-Model} The packet
arrival $N_{l}(t)$ is a Poisson process with intensity $\lambda_{l}$.
Specifically, $N_{l}(t)$ follows a probability law given by, 
\[
\mbox{Pr}\left(N_{l}(t+dt)-N_{l}(t)=1\big|N_{l}(t)\right)=\lambda_{l}dt.
\]
 
\end{asmQED}

The queueing dynamics of the wireless adhoc network can be described
by the following SDE,
\begin{equation}
dq_{l}=-\mu_{l}dt+dN_{l}.\label{eq:sys-QSI-model}
\end{equation}
The first term in (\ref{eq:sys-QSI-model}) corresponds to the packet
departure and the second term corresponds to the random packet arrival.
Using Little's Law \cite{Little1961:law}, the average delay of the
$l$-th link ($l$-th flow) is given by $\overline{T}_{l}=\overline{q}_{l}/\lambda_{l}$,
where $\overline{q}_{l}$ is the average backlog for the $l$-th queue.
As a result, there is no loss of generality to study the average queue
length $\overline{q}_{l}$ as this is proportional to the average
delay. Obviously, the average queue length (or average delay) of the
wireless adhoc network depends on how we allocate the transmit power
$p_{l}(t)$ and data rate $\mu_{l}(t)$ of each link in the network.
In the next section, we shall briefly review the MWQ algorithm, which
is known to be a \emph{throughput optimal control} (in queue stability
sense).

\subsection{Queue Stability and Max-Weighted Queue (MWQ) Algorithm\label{sub:Max-Weighted-Queue-(MWQ)}}

There are different ways to control the power $p_{l}(t)$ and rates
$\mu_{l}(t)$ of the wireless networks but a reasonable algorithm
(in delay sense) should adapt to both the CSI (to capture good transmission
opportunity) and the QSI (to capture the urgency). In particular,
we are interested in control policy that achieves a maximum queue
stability region. We now first define the notion of \emph{queue stability},
\emph{stability region} and \emph{throughput optimal control}.
\begin{lyxDefQED}
[Queueing Stability] A queue is called \emph{stable} if $\lim\sup_{t\to\infty}\frac{1}{t}\int_{0}^{t}\mathbb{E}\left[\|\mathbf{q}(\tau)\|\right]<\infty$.
\end{lyxDefQED}

The \emph{stability region} $\overline{\mathcal{C}}$ is defined as
the closure of the set of all the arrival rate vectors $\left\{ \lambda_{l}\right\} $
that can be stabilized under some control algorithm that conforms
to the power constraint $\mathbb{E}\left[\mathbf{p}\right]\in\mathcal{P}$
\cite{Georgiadis2006}. A control policy that is \emph{throughput
optimal} is characterized in the sense that it stabilizes all the
arrival rate vectors $\left\{ \lambda_{l}\right\} $ within the stability
region $\overline{\mathcal{C}}$ \cite{Tassiulas1992}. The throughput
optimal policy is not unique and there are various known methods to
achieve the maximum queue stability region. For technical reasons,
we define a convex compact domain $\mathcal{P}=\{p:0\leq p\leq2^{L\lambda_{\max}}/h_{0}^{2}\}$.
Using Lyapunov techniques, a \emph{throughput optimal} (in stability
sense) formulation for the power and rate control actions at each
time slot $t$ is given in the following. 
\begin{problem}
[MWQ Formulation]

\begin{eqnarray}
\max_{\mathbf{p}\in\mathcal{P},\bm{\mu}\succeq\mathbf{0}} & \sum_{l}\left[q_{l}(t)\mu_{l}(t)-Vp_{l}(t)\right] & \text{}\label{eq:prob-weighted-queue}\\
\mbox{subject to} & \bm{\mu}(t)=(\mu_{l}(t),\dots,\mu_{l}(t))^{T}\in\mathcal{C}(\mathbf{p}(t),\mathbf{h}(t))\label{eq:prob-const-capacity-region}
\end{eqnarray}
where the physical layer capacity region $\mathcal{C}(\mathbf{p}(t),\mathbf{h}(t))$
is a polyhedron defined by the constraints in (\ref{eq:capacity-region})
for all receiving nodes $n$. The parameter $V$ acts as a Lagrange
multiplier which controls the tradeoff between the average delay and
the average power of the wireless network. ~\hfill\IEEEQEDclosed
\end{problem}

Note that the MWQ optimization problem in (\ref{eq:prob-weighted-queue})-(\ref{eq:prob-const-capacity-region})
is parameterized by the current CSI $\mathbf{h}(t)=\left\{ h_{1}(t),\dots,h_{L}(t)\right\} $
and the QSI $\mathbf{q}(t)=\{q_{1}(t)\dots q_{L}(t)\}$. As a result,
the optimal solution $\mathbf{p}^{*}(\mathbf{h}(t),\mathbf{q}(t))$
and $\bm{\mathbf{\mu}}^{*}(\mathbf{h}(t),\mathbf{q}(t))$ of the MWQ
problem is also parameterized by the CSI and QSI $(\mathbf{h}(t),\mathbf{q}(t))$. 

Due to the interference coupling in the MWQ problem, there are no
closed form solutions for $\mathbf{p}^{*}(\mathbf{h},\mathbf{q})$
and $\bm{\mu}^{*}(\mathbf{h},\mathbf{q})$ despite the problem in
(\ref{eq:prob-weighted-queue})-(\ref{eq:prob-const-capacity-region})
being convex. To solve the MWQ problem in (\ref{eq:prob-weighted-queue})-(\ref{eq:prob-const-capacity-region}),
we first have the following lemma regarding the rate allocation $\bm{\hat{\mu}}(\mathbf{p};\mathbf{h},\mathbf{q})$
given the power. 
\begin{lyxLemQED}
[Optimal rate allocation \cite{Tse1998}]\label{lem:Optimal-rate-allocation}
Let $\bm{\pi}=\{\pi(1),\pi(2),\lyxmathsym{\ldots}\pi(L)\}$ be a permutation
of the flow indices sorted in descendent order of the QSI $q_{l}$,
i.e. $q_{\pi(1)}\geq q_{\pi(2)}\geq\dots\geq q_{\pi(L)}$. Given a
power allocation \textbf{$\mathbf{p}=[p_{1},p_{2},\dots,p_{L}]$}
, the optimal rate allocation solution of the MWQ problem in (\ref{eq:prob-weighted-queue})-(\ref{eq:prob-const-capacity-region})
is given by
\begin{eqnarray}
\hat{\mu}_{\pi(1)} & = & \log\left(1+\left|h_{\pi(1)}\right|^{2}p_{\pi(1)}\right)\label{eq:mu-hat-1}\\
\hat{\mu}_{\pi(k)} & = & \log\left(1+\sum_{i=1}^{k}\left|h_{\pi(i)}\right|^{2}p_{\pi(i)}\right)-\log\left(1+\sum_{i=1}^{k-1}\left|h_{\pi(i)}\right|^{2}p_{\pi(i)}\right),\quad k=2,\dots,L\label{eq:mu-hat-k}
\end{eqnarray}

\end{lyxLemQED}

Intuitively, given a power allocation, the optimal rate allocation
vector $\hat{\bm{\mu}}=\{\hat{\mu}_{1},\dots,\hat{\mu}_{L}\}$ is
given by one of the vertices of the polyhedron $\mathcal{C}(\mathbf{p},\mathbf{h})$.
In addition, the vertices are achieved by the SIC with decoding order
$\bm{\pi}$. As a result, finding the optimal $\hat{\bm{\mu}}(\mathbf{p};\mathbf{h},\mathbf{q})$
is equivalent to a linear programming problem, which requires $L$
steps of iterations. Hence, we can focus on the power optimization
in the MWQ problem given by  
\begin{equation}
\max_{\mathbf{p}(t)\in\mathcal{P}}\quad\mathcal{L}(\mathbf{p}(t);\mathbf{h}(t),\mathbf{q}(t))=\sum_{l=1}^{L}q_{l}(t)\hat{\mu}_{l}(\mathbf{p}(t);\mathbf{h}(t),\mathbf{q}(t))-V\sum_{l=1}^{L}p_{l}(t).\label{eq:prob-power-allocation}
\end{equation}
Using an iterative projected gradient search algorithm to find the
optimal solution in (\ref{eq:prob-power-allocation}), we derive the
following power control algorithms dynamics \cite{Arrow1958}, 
\begin{equation}
\dot{\mathbf{p}}=\kappa\left[\nabla\mathcal{L}\left(\mathbf{p};\mathbf{h}(t),\mathbf{q}(t)\right)\right]_{\mathbf{p}}^{\mathcal{P}}\label{eq:alg-pwr-update-0}
\end{equation}
where $\kappa$ is a step size parameter, and the entry-wide projection
operator $\left[\centerdot\right]_{z}^{\mathcal{P}}$ is defined as
$[x]_{z}^{\mathcal{P}}:=0$, if $z\in\partial\mathcal{P}$ is on the
boundary of $\mathcal{P}$ and $z+xdt\in\mathcal{P}$, and $[x]_{z}^{\mathcal{P}}:=x$,
otherwise. Hence, the queue dynamics of the wireless adhoc network
under MWQ control is determined by the following \emph{coupled SDEs}.
\begin{eqnarray}
dp_{l} & = & \kappa\left[\frac{\partial}{\partial p_{l}}\mathcal{L}\left(\mathbf{p};\mathbf{h},\mathbf{q}\right)\right]_{p_{l}}^{\mathcal{P}}\label{SDE:system-dynamic-0-p}\\
dh_{l} & = & -\frac{1}{2}a_{l}h_{l}dt+a_{l}^{\frac{1}{2}}dw_{l}+dv_{l}\label{SDE:system-dynamic-0-h}\\
dq_{l} & = & -\hat{\mu}_{l}(\mathbf{p};\mathbf{h},\mathbf{q})dt+dN_{l},\quad\forall l=1,\dots,L.\label{SDE:system-dynamic-0-q}
\end{eqnarray}

In existing works, the convergence of the gradient algorithm in (\ref{eq:alg-pwr-update-0})
and the throughput optimality of the MWQ in (\ref{eq:prob-weighted-queue})
are all based on an important assumption that the CSI and the QSI
$(\mathbf{h},\mathbf{q})$ remains constant during the algorithm iterations
in (\ref{eq:alg-pwr-update-0}). However, in practice, this may not
be satisfied especially for fast fading channels and heavy traffic
arrivals. In this paper, we are interested in studying the convergence
behavior as well as the throughput and delay penalty of the iterative
MWQ algorithm when the CSI and the QSI are changing at a similar timescale
as that of the MWQ iterations.

\subsection{Virtual Stochastic Dynamic Systems\label{sub:VSDS-Virtual-Stochastic-Dynamic}}

In this subsection, we show that studying the convergence behavior
of MWQ algorithm iterations in (\ref{eq:alg-pwr-update-0}) and the
queue stability can be transformed into an equivalent problem of \emph{stochastic
stability} in a \emph{virtual stochastic dynamic system} (VSDS). As
a result of this association, we can focus on analyzing the behavior
of the VSDS instead of the original complicated MWQ algorithm dynamics.
We first have a few definitions. 
\begin{lyxDefQED}
[Equilibrium Point]\label{def:Equilibrium-Point} Given the CSI
and QSI parameter $(\mathbf{h},\mathbf{q})$, $\mathbf{p}^{*}(\mathbf{h},\mathbf{q})$
is called an equilibrium point of the MWQ algorithm dynamics in (\ref{eq:alg-pwr-update-0})
if $\nabla\mathcal{L}\left(\mathbf{p}^{*};\mathbf{h},\mathbf{q}\right)=0$.
\end{lyxDefQED}

\begin{figure}
\begin{centering}
\includegraphics[scale=0.6]{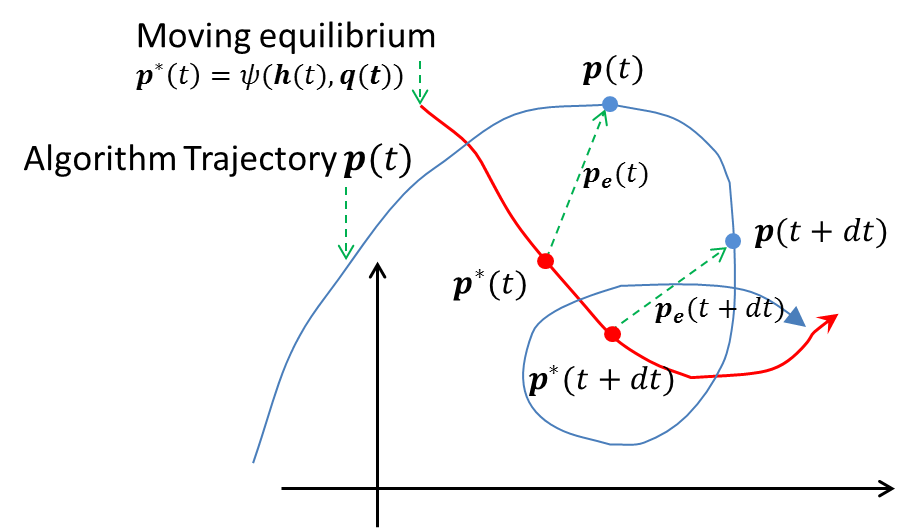}
\par\end{centering}

\caption{\label{fig:tracking-moving-eqiulibrium}An illustration of the algorithm
trajectory for solving an MWQ problem with time-varying CSI and QSI.
The dynamics of the CSI and QSI excite the equilibrium $\mathbf{p}^{*}(t)$
to move around, and hence the convergence of $\mathbf{p}(t)$ is not
guaranteed.}
\end{figure}

When the CSI and QSI $(\mathbf{h},\mathbf{q})$ are quasi-static,
the equilibrium point $\mathbf{p}^{*}(\mathbf{h},\mathbf{q})$ is
fixed and it has been shown \cite{Arrow1958,Feijer2010} that the
MWQ algorithm iterations in (10) converges to $\mathbf{p}^{*}(\mathbf{h},\mathbf{q})$
after sufficient iterations. However, when $(\mathbf{h},\mathbf{q})$
are time-varying, the equilibrium point $\mathbf{p}^{*}(\mathbf{h},\mathbf{q})$
is also time-varying as illustrated in Fig. \ref{fig:tracking-moving-eqiulibrium}
and it is not known if the MWQ iterations can track the \emph{moving
target}. To measure the tracking performance, we define the \emph{tracking
error vector} between the MWQ algorithm trajectory and the moving
equilibrium point as below.

\begin{lyxDefQED}
[Tracking Error Vector]\label{def:Tracking-Error} The tracking
error vector of the MWQ algorithm is a vector difference between the
algorithm trajectory $\mathbf{p}(t)$ and the \emph{target} equilibrium
point $\mathbf{p}^{*}(\mathbf{h}(t),\mathbf{q}(t))$, i.e., $\mathbf{p}_{e}(t)=\mathbf{p}(t)-\mathbf{p}^{*}(t)$.
\end{lyxDefQED}

For a notation convenience, let $\psi:(\mathbf{h},\mathbf{q})\mapsto\mathbf{p}^{*}(\mathbf{h},\mathbf{q})$
be a mapping from the current CSI and QSI $(\mathbf{h},\mathbf{q})$
to the equilibrium point $\mathbf{p}^{*}(\mathbf{h},\mathbf{q})$.
From Definition \ref{def:Tracking-Error}, the drift of the tracking
error can be expressed as 
\begin{eqnarray}
d\mathbf{p}_{e} & = & d\mathbf{p}-d\mathbf{p}^{*}\label{eq:dpe}\\
 & = & \kappa\left[\nabla\mathcal{L}\left(\mathbf{p};\mathbf{h},\mathbf{q}\right)\right]_{\mathbf{p}}^{\mathcal{P}}dt-\psi_{q}(\mathbf{h},\mathbf{q})d\mathbf{q}-\psi_{h}(\mathbf{h},\mathbf{q})d\mathbf{h}\nonumber 
\end{eqnarray}
where $\psi_{q}(\centerdot)=\frac{\partial}{\partial\mathbf{q}}\psi(\mathbf{h},\mathbf{q})$
and $\psi_{h}(\centerdot)=\frac{\partial}{\partial\mathbf{h}}\psi(\mathbf{h},\mathbf{q})$
are partial derivatives of the equilibrium point $\mathbf{p}^{*}=\psi(\mathbf{h},\mathbf{q})$
over the current QSI $\mathbf{q}$ and CSI $\mathbf{h}$. They represent
the sensitivity of $\mathbf{p}^{*}(\mathbf{h},\mathbf{q})$ with respect
to the variations of the CSI and QSI $(\mathbf{h},\mathbf{q})$. The
terms $\psi_{q}(\mathbf{h},\mathbf{q})d\mathbf{q}$ and $\psi_{h}(\mathbf{h},\mathbf{q})d\mathbf{h}$
represent the change of the optimal power $d\mathbf{p}^{*}$ corresponding
to the time-varying QSI $d\mathbf{q}(t)$ and CSI $d\mathbf{h}(t)$,
respectively. Note that, as $\mathbf{h}$ is complex, $\psi_{h}(\mathbf{h},\mathbf{q})d\mathbf{h}$
is defined as $\psi_{h_{l}}dh_{l}=\frac{\partial\psi}{\partial x_{l}}dx_{l}+\frac{\partial\psi}{\partial y_{l}}dy_{l}$,
for each complex component%
\footnote{The complex derivative for a real value function $\psi(h)$ is defined
as $\frac{\partial\psi}{\partial h}=\frac{1}{2}\left(\frac{\partial\psi}{\partial x}-i\frac{\partial\psi}{\partial y}\right)$
and $\frac{\partial\psi}{\partial\overline{h}}=\frac{1}{2}\left(\frac{\partial\psi}{\partial x}+i\frac{\partial\psi}{\partial y}\right)$,
for $h=x+iy$. The Taylor expansion of $\psi(h)$ is thus given by
$d\psi=\frac{\partial\psi}{\partial h}dh+\frac{\partial\psi}{\partial\overline{h}}d\overline{h}=\frac{\partial\psi}{\partial x}dx+\frac{\partial\psi}{\partial y}dy$
\cite{Brandwood1983}.%
} $h_{l}=x_{l}+iy_{l}$. Taking $\mathbf{p}=\mathbf{p}^{*}+\mathbf{p}_{e}$,
we denote 
\[
f(\mathbf{p}_{e};\mathbf{h},\mathbf{q})\triangleq\kappa\left[\nabla\mathcal{L}\left(\mathbf{p}_{e}+\mathbf{p}^{*};\mathbf{h},\mathbf{q}\right)\right]_{\mathbf{p}_{e}+\mathbf{p}^{*}}^{\mathcal{P}}
\]
as a mapping of the gradient iterations. Using the system dynamics
of $\mathbf{h}(t)$ and $\mathbf{q}(t)$ in (\ref{SDE:system-dynamic-0-h})-(\ref{SDE:system-dynamic-0-q}),
we construct a \emph{stochastic error dynamic system} to describe
the tracking error process $\mathbf{p}_{e}(t)$ as follows. 
\begin{lyxDefQED}
[Stochastic Error Dynamic System (SEDS)] The stochastic error dynamic
system is characterized by the following SDE 
\begin{equation}
d\mathbf{p}_{e}=f_{e}(\mathbf{p}_{e};\mathbf{h},\mathbf{q})dt+b_{e}(\mathbf{p}_{e};\mathbf{h},\mathbf{q})d\mathbf{N}(t)+c_{e}(\mathbf{p}_{e};d\mathbf{W}(t),d\mathbf{V}(t))\label{sde:power-error-dynamic}
\end{equation}
where $f_{e}(\mathbf{p}_{e};\mathbf{h},\mathbf{q})=f(\mathbf{p}_{e};\mathbf{h},\mathbf{q})-\psi_{q}(\centerdot)\hat{\bm{\mu}}(\centerdot)+\frac{1}{2}\psi_{h}(\centerdot)A\mathbf{h}$,
$b_{e}(\centerdot)=-\psi_{q}(\centerdot)$, and \\$c_{e}(\centerdot)=-\psi_{h}(\centerdot)(A^{\frac{1}{2}}d\mathbf{W}(t)+d\mathbf{V}(t))$.
$A=\mbox{diag}\{a_{1},\dots,a_{L}\}$ is a matrix of CSI correlation
coefficient in (\ref{sde:CSI-model}).
\end{lyxDefQED}

It is known that when the CSI and QSI $(\mathbf{h},\mathbf{q})$ are
static, the MWQ algorithm trajectory always converges to the static
equilibrium point $\mathbf{p}^{*}(\mathbf{h},\mathbf{q})$. However,
when the CSI and QSI are time-varying in a similar timescale as the
MWQ algorithm iterations, the algorithm convergence is not obvious.
To study the behavior of the algorithm dynamics induced by the time-varying
CSI and QSI, we construct a Virtual Stochastic Dynamic System (VSDS),
which combines the overall dynamics of the CSI and QSI in (\ref{SDE:system-dynamic-0-h})
and (\ref{SDE:system-dynamic-0-q}) with the Stochastic Error Dynamic
System (SEDS) in (\ref{sde:power-error-dynamic}) as follows. 
\begin{lyxDefQED}
[Virtual Stochastic Dynamic System (VSDS)] Let $\mathbf{z}=(\mathbf{p}_{e},\mathbf{h},\mathbf{q})$
be a joint system state. The virtual stochastic dynamic system is
characterized by the following coupled SDE, 
\begin{equation}
\mathcal{Z}:\qquad d\mathbf{z}=F(\mathbf{z})dt+B(\mathbf{z})d\mathbf{N}+C(\mathbf{z},d\mathbf{W},d\mathbf{V})\label{sde:VSDS}
\end{equation}
where 
\[
F(\mathbf{z})=\left[\begin{array}{c}
f(\mathbf{p}_{e};\mathbf{h},\mathbf{q})-\psi_{q}\hat{\bm{\mu}}(\mathbf{p}_{e}+\mathbf{p}^{*})+\frac{1}{2}\psi_{h}A\mathbf{h}\\
-\frac{1}{2}A\mathbf{h}\\
-\hat{\bm{\mu}}(\mathbf{p}_{e}+\mathbf{p}^{*})
\end{array}\right]
\]
\[
B(\mathbf{z})=\left[\begin{array}{c}
-\psi_{q}(\mathbf{h},\mathbf{q})\\
\mathbf{0}_{2L\times L}\\
\mathbf{I}_{L}
\end{array}\right]\quad\mbox{and}\quad C(\mathbf{z},d\mathbf{W}(t))=\left[\begin{array}{c}
-\psi_{h}(\mathbf{h},\mathbf{q})(A^{\frac{1}{2}}d\mathbf{W}+d\mathbf{V})\\
A^{\frac{1}{2}}d\mathbf{W}+d\mathbf{V}\\
\mathbf{0}_{L}
\end{array}\right].
\]

\end{lyxDefQED}

\begin{figure}
\begin{centering}
\includegraphics[scale=0.3]{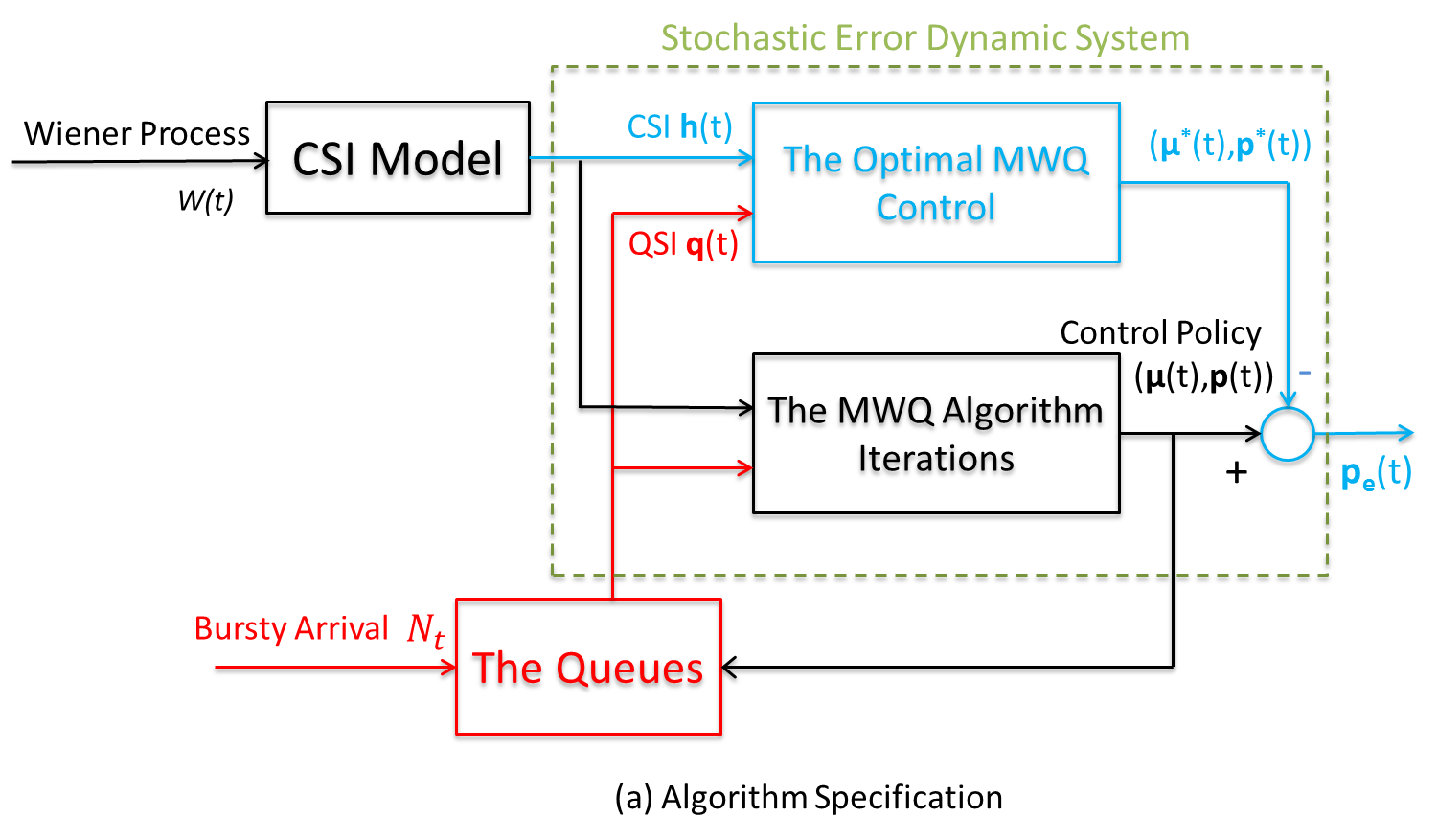} \includegraphics[scale=0.3]{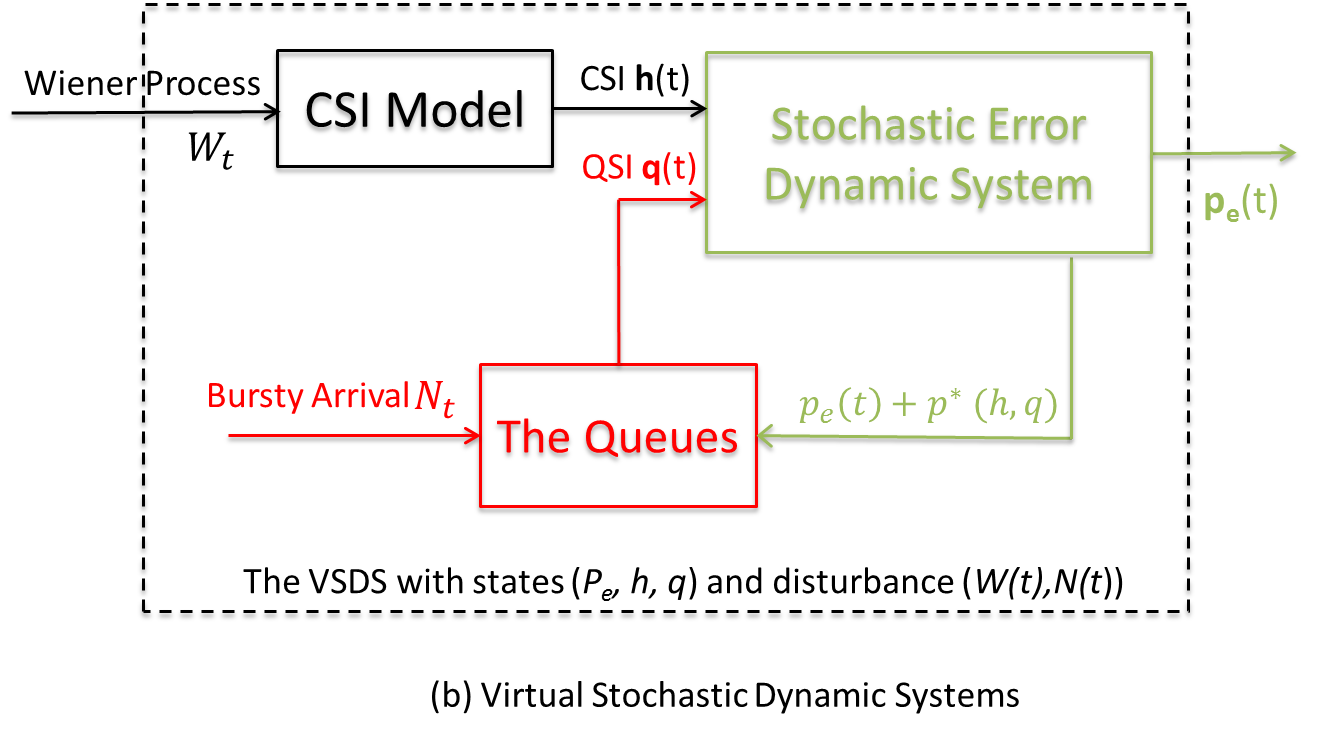}
\par\end{centering}

\caption{\label{fig:The-virtual-stochastic-VSDS}An illustration of the connection
between the MWQ algorithm dynamics and Virtual Stochastic Dynamic
System (VSDS). Fig. (a) illustrates the dynamics in the MWQ algorithm
domain. The control policies $(\bm{\mu}(t),\mathbf{p}(t))$ from the
MWQ algorithm iterations are driven by the CSI dynamics $\mathbf{h}(t$)
and the QSI dynamics $\mathbf{q}(t)$. Fig. (b) illustrates the coupled
MWQ iterations, CSI and QSI from the VSDS perspective, where the power
tracking error $\mathbf{p}_{e}(t)$, the CSI and QSI $(\mathbf{h}(t),\mathbf{q}(t))$
are modeled as a joint state of the SDE, which is driven by external
stochastic processes $W_{t}$ and $N_{t}$. }
\end{figure}

Fig. \ref{fig:The-virtual-stochastic-VSDS} illustrates the inter-connection
between the key components in the VSDS. Fig. \ref{fig:The-virtual-stochastic-VSDS}(a)
illustrates the dynamics in the MWQ algorithm domain. Specifically
the queueing dynamics $\mathbf{q}(t)$ is driven by the bursty arrival
process $\mathbf{N}(t)$ as well as the control policy $(\bm{\mu}(t),\mathbf{p}(t))$.
At the same time, the control actions $(\bm{\mu}(t),\mathbf{p}(t))$
are driven by the MWQ algorithm iterations, which depend on the CSI
$\mathbf{h}(t$) and the QSI $\mathbf{q}(t)$. Fig. \ref{fig:The-virtual-stochastic-VSDS}(b)
illustrates the dynamics in the VSDS domain. The system consists of
the SEDS (driving the tracking error process $\mathbf{p}_{e}(t)=\mathbf{p}(t)-\mathbf{p}^{*}(t)$)
as well as the CSI $\mathbf{h}(t)$ and QSI $\mathbf{q}(t)$ driven
by external processes $\mathbf{W}(t)$ and $\mathbf{N}(t)$. 

We show in the following theorem that, studying the convergence behavior
of the MWQ algorithm (\ref{eq:alg-pwr-update-0}) is the same as studying
the stability property of the VSDS in (\ref{sde:VSDS}). Also, evaluating
the stability of queue backlogs driven by the MWQ algorithm dynamics
is equivalent to investigating the stability property of the system
state $\mathbf{z}(t)$ in the VSDS.
\begin{lyxThmQED}
[Connections between the MWQ Algorithm Dynamics and the VSDS]\label{Thm: Connections-virtual-algorithm}
The actual queue trajectory of the MWQ algorithm in (\ref{SDE:system-dynamic-0-q})
is the same as the solution process $\mathbf{q}(t)$ in the VSDS in
(\ref{sde:VSDS}).  Furthermore, the power control trajectory of the
MWQ algorithm in (\ref{SDE:system-dynamic-0-p}) converges to the
equilibrium $\mathbf{p}^{*}(\mathbf{h},\mathbf{q})$ if and only if
the SDE in (\ref{sde:power-error-dynamic}) is globally asymptotically
stable at $\mathbf{p}_{e}=\mathbf{0}$, i.e., given any initial state
$\mathbf{p}_{e}(0)\in\mathbb{R}_{+}^{L}$, $\lim_{t\to\infty}\mbox{Pr}\left(\mathbf{p}_{e}(t)=0\right)=1$. 
\end{lyxThmQED}
\begin{proof}
Please refer to Appendix \ref{app:thm1-connection} for the proof.
\end{proof}

As a result of Theorem \ref{Thm: Connections-virtual-algorithm},
we can focus on the VSDS dynamics in order to study the delay performance
penalty of MWQ due to time varying CSI and QSI. Nevertheless, due
to the mutual coupling of the SDEs in the VSDS, it is still difficult
to study its stability behavior. In the rest of the paper, we will
focus on extending the stochastic Foster-Lyapunov method \cite{Meyn93:StabilityIII-Foster}
to derive the stochastic stability results of the VSDS.

\section{Performance Analysis of MWQ Algorithm under Time-Varying Arrivals
and Channels}

In this section, we shall analyze the tracking performance of the
MWQ algorithm under time-varying channels. We bridge the connection
between the property of the Lyapunov stochastic drift and the stochastic
stability of the corresponding VSDS. Following this result, we then
derive an expected queue bound under the MWQ algorithm in time-varying
channels.

\subsection{Stochastic Stability of Random Process}

Let $\mathbf{z}=(\mathbf{p}_{e},\mathbf{h},\mathbf{q})$ be a joint
state of the VSDS in (\ref{sde:VSDS}), where $\mathbf{p}_{e}$ is
the tracking error, $\mathbf{h}$ is the channel coefficient and $\mathbf{q}$
is the queue backlog. Denote $\mathbf{z}(t)$ as the stochastic process
starting from $t=0$ with initial state $\mathbf{z}(0)$. We have
the following definition of stochastic stability to characterize the
behavior of $\mathbf{z}(t)$.
\begin{lyxDefQED}
[Stochastic Stability]\label{def:Stochastic-Stability-1} Given
any initial state $\mathbf{z}(0)\in\mathcal{Z}$, the stochastic process
$\mathbf{z}(t)$ is globally stochastically stable, if there exists
$0\leq D<\infty$, such that 
\[
\lim\sup_{t\to\infty}\frac{1}{t}\int_{0}^{t}\mathbb{E}\left\Vert \mathbf{z}(t)\right\Vert \leq D.
\]

\end{lyxDefQED}

Notice that this definition is analogue to the usual concept of \emph{stability}
in \emph{deterministic }system \cite{Khalil1996}, whereas, the condition
here is taken over a time averaged expectation. This general criteria
can be applied to a non-stationary process, such as a queueing system
with different classes of services. In fact, in this work, we do not
require the queue dynamics and the MWQ algorithm trajectory to be
stationary.

Define a Lyapunov function of the state $\mathbf{z}(t)$ as $V(\mathbf{z})=\mathbf{z}^{H}\mathbf{z}$.
We can investigate the evolution of the Lyapunov function by studying
its drift along the state trajectory. Analogue to the discrete-time
one-step \emph{conditional Lyapunov drift} in \cite{Georgiadis2006},
we define the continuous time Lyapunov drift generator as 
\begin{equation}
LV(\mathbf{z}(t))=\lim_{\delta\downarrow0}\frac{\mathbb{E}\left[V(\mathbf{z}(t+\delta))-V(\mathbf{z}(t))|\mathbf{z}(t)\right]}{\delta}\label{eq:infinitesimal-1}
\end{equation}
where the expectation (conditioned on the current state $\mathbf{z}(t)$)
is taken over the randomness of the CSI and the arrival to the QSI.
The Lyapunov drift represents the expected evolving direction of the
Lyapunov function $V(\mathbf{z})$ from the current state $\mathbf{z}(t)$,
and $LV(\mathbf{z})$ is called an \emph{infinitesimal estimator}
of $V(\mathbf{z})$. As $V(\mathbf{z})$ is a norm-like function \cite{Khalil1996},
the boundedness of the Lyapunov function implies the boundedness of
state $\mathbf{z}(t)$ and the dynamics of the Lyapunov function reveals
the evolution of state $\mathbf{z}(t)$. For example, when the drift
is negative, $\|\mathbf{z}(t)\|$ is most likely decreasing. The  Lyapunov
drift $LV(\mathbf{z})$ can also be derived from the SDE of $\mathbf{z}(t)$
as stated in the following lemma.
\begin{lyxLemQED}
[Continuous-Time Lyapunov Drift \cite{Kushner:Stoch-contrl:1967}]\label{lem:Lyapunov-Drift-of-SDE-1}
Suppose that there is a $d$-dimensional stochastic process $\mathbf{z}(t)$
described by a SDE 
\[
d\mathbf{z}=f(\mathbf{z})dt+g(\mathbf{z})d\mathbf{W}+h(\mathbf{z})d\mathbf{N}
\]
where $\mathbf{W}(t)\in\mathbb{C}^{L}$ is a standard complex Wiener
process and $\mathbf{N}(t)=(N_{1}(t),\dots,N_{L}(t))\in\mathbb{Z}_{+}^{L}$
is a Poisson process with intensities $\lambda_{l}$, $l=1,\dots,L$.
For any  given Lyapunov function $V(\mathbf{z})\in\mathcal{C}^{2}:\mathcal{Z}\to\mathbb{R}_{+}$
that has compact support \cite{Kushner:Stoch-contrl:1967}, the stochastic
Lyapunov drift can be written as 
\begin{eqnarray*}
LV(\mathbf{z}) & = & 2\frac{\partial V(\mathbf{z})}{\partial\mathbf{z}}f(\mathbf{z})+\frac{1}{2}\mbox{tr}\left[g(\mathbf{z})^{H}\frac{\partial^{2}V(\mathbf{z})}{\partial\overline{\mathbf{z}}\partial\mathbf{z}}g(\mathbf{z})+g(\mathbf{z})\frac{\partial^{2}V(\mathbf{z})}{\partial\mathbf{z}\partial\overline{\mathbf{z}}}g(\mathbf{z})^{H}\right]\\
 &  & \qquad+\sum_{l=1}^{L_{2}}\lambda_{l}\left(V(\mathbf{z}+h^{(l)}(\mathbf{z}))-V(\mathbf{z})\right)
\end{eqnarray*}
where $h^{(l)}(\mathbf{z})$ is the $l$-th column of $h(\mathbf{z})$. 
\end{lyxLemQED}

The above lemma establishes a connection between the infinitesimal
estimator $LV(\mathbf{z})$ and the specific SDE. The proof is similar
to that in  \cite{Kushner:Stoch-contrl:1967} with a notation extension
to complex variables \cite{Brandwood1983}. By exploiting the property
of the Lyapunov drift, we can characterize the stochastic stability
of random process $\mathbf{z}(t)$ described by the SDE in (\ref{sde:VSDS}).
We summarize the result in the following theorem.
\begin{lyxThmQED}
[Stochastic Stability from Lyapunov Drift]\label{thm:stoch-stability-region-z}
Suppose the stochastic Lyapunov drift of the process $\mathbf{z}(t)$
satisfies 
\begin{equation}
LV(\mathbf{z})\leq-a\|\mathbf{z}\|+g(\mathbf{s})\label{eq:Lya-criteria-property}
\end{equation}
for all $\mathbf{z}\in\mathcal{Z}$, where $a$ is some positive constant
and $\mathbf{s}(t)$ is a stochastic process that satisfies 
\[
\lim\sup_{t\to\infty}\frac{1}{t}\int_{0}^{t}\mathbb{E}\left[g(\mathbf{s}(\tau))\right]d\tau\leq d
\]
for some function $g:\mathbf{s}\mapsto\mathbb{R}$ and $d<\infty$.
Then the process $\mathbf{z}(t)$ is stochastically stable, and 
\[
\lim\sup_{t\to\infty}\frac{1}{t}\int_{0}^{t}\mathbb{E}\|\mathbf{z}(\tau)\|d\tau\leq\frac{d}{a}.
\]

\end{lyxThmQED}
\begin{proof}
Please refer to Appendix \ref{app:lem-stochastic-stability-region-z}
for the proof.
\end{proof}

The above result is an  extension of the Foster-Lyapunov criteria
for continuous time processes in \cite{Meyn93:StabilityIII-Foster}.
If $\mathbf{z}(t)$ is a system state that relates to the queue length
and the tracking error of the power variable, the stochastic stability
forms an estimation on the queue bound as well as the power penalty
due to the time-varying parameters. The advantage of the Foster-Lyapunov
method enables a qualitative analysis of the VSDS without explicitly
solving the SDE. In the following, we shall illustrate how to construct
a Lyapunov drift for  the VSDS.

\subsection{Stability Analysis of the MWQ Algorithm\label{sub:Performance-Analysis-MWQ}}

In this section, we shall apply the stochastic stability analysis
method to study the stability of the VSDS (\ref{sde:VSDS}). Specifically,
according to Lemma \ref{lem:Lyapunov-Drift-of-SDE-1}, the Lyapunov
drift for the VSDS (\ref{sde:VSDS}) is given by%
\footnote{Note that Lemma \ref{lem:Lyapunov-Drift-of-SDE-1} does not specify
the drift for the term $d\mathbf{V}$. However, as the reflection
$|d\mathbf{V}|\leq|A^{\frac{1}{2}}d\mathbf{W}|$, we can treat $\mathbf{V}(t)$
as $A^{\frac{1}{2}}\mathbf{W}(t)$ and yield an upper bound for $LV(\mathbf{z})$.%
} 
\begin{eqnarray}
LV(\mathbf{z}) & = & 2\mathbf{z}^{H}F(\mathbf{z})+\frac{1}{2}\mbox{tr}\left[2C(\mathbf{z},\centerdot)^{H}C(\mathbf{z},\centerdot)\right]+\sum_{l=1}^{L}\lambda_{l}\left[V(\mathbf{z}+B^{(l)})-V(\mathbf{z})\right]\nonumber \\
 & \leq & 2\mathbf{p}_{e}^{T}f(\mathbf{p}_{e};\mathbf{h},\mathbf{q})-2\mathbf{p}_{e}^{T}\psi_{q}(\mathbf{h},\mathbf{q})\hat{\bm{\mu}}(\mathbf{p}_{e}+\mathbf{p}^{*})+\mathbf{p}_{e}^{T}\psi_{h}(\mathbf{h},\mathbf{q})A\mathbf{h}\nonumber \\
 &  & \quad-\mathbf{h}^{H}A\mathbf{h}-2\mathbf{p}_{e}^{T}\psi_{q}(\mathbf{h},\mathbf{q})\bm{\lambda}-2\sum_{l=1}^{L}q_{l}\left[\hat{\mu}_{l}(\mathbf{p}_{e}+\mathbf{p}^{*})-\lambda_{l}\right]\nonumber \\
 &  & \quad+\mbox{tr}\left(2\left(A^{\frac{1}{2}}\right)^{T}\psi_{h}^{H}\psi_{h}A^{\frac{1}{2}}+\psi_{q}^{T}\psi_{q}\right)+\sum_{l=1}^{L}\left(2a_{l}+\lambda_{l}\right)\label{eq:LV-VSDS}
\end{eqnarray}
where $B^{(l)}$ stands for the $l$-th column of $B(\mathbf{z})$
and $\Lambda=\mbox{diag}\left\{ \lambda_{1},\dots,\lambda_{L}\right\} $
is the arrival matrix.

As illustrated in Theorem \ref{thm:stoch-stability-region-z}, negative
drift terms in $LV(\mathbf{z})$ are desirable because they can drive
the stochastic state process towards the origin and contribute to
stabilization of the VSDS. In the following, we shall analyze the
key terms on the R.H.S. of (\ref{eq:LV-VSDS}) and discuss their contributions.
Intuitively, a fast MWQ algorithm and a high transmission rate can
contribute to driving the $LV(\mathbf{z})$ negative and we shall
elaborate on such properties in the following. 

We first define the following, which will be used throughout the analysis.
Let $S_{\min}(t)=\min_{l}\left\{ \left|h_{l}(t)\right|^{2}\right\} $
and $S_{\max}(t)=\max_{l}\left\{ \left|h_{l}(t)\right|^{2}\right\} $
be the minimum and maximum channel gains among $L$ transmission links
at time $t$, respectively. Notice that $\left|h_{l}\right|^{2}$
has stationary distributions and we denote its cumulative distribution
function as $F_{h}(x)$. Thus $S_{\min}$ and $S_{\max}$ are also
ergodic processes with stationary distributions given by the $L$-th
\emph{order statistics }as $F_{S}^{\min}(s)=\mbox{P}(S_{\min}\leq s)=1-\left[1-\mbox{P}\left(|h|^{2}\leq s\right)\right]^{L}=1-\left[1-F_{h}(s)\right]^{L}$
and $F_{S}^{\max}(s)=\mbox{P}\left(S_{\max}\leq s\right)=\mbox{P}\left(|h|^{2}\leq s\right)^{L}=F_{h}(s)^{L}$.

The following lemma summarizes the contribution of the convergence
speed of the MWQ iterations in (\ref{eq:alg-pwr-update-0}) to the
drift $LV(\mathbf{z})$ in (\ref{eq:LV-VSDS}). 
\begin{lyxLemQED}
[Negative Drift Contribution of Convergence Speed in MWQ Gradient
Iteration]\label{lem:conv-speed-gradient-alg} Given any CSI and
QSI realizations $\mathbf{h}(t)=\mathbf{h}$ and $\mathbf{q}(t)=\mathbf{q}\succ\mathbf{1}$,
there exists $\alpha(S_{\min},S_{\max})>0$ that satisfies%
\footnote{For vectors $\mathbf{a}=(a_{1},a_{2},\dots,a_{L})$ and $\mathbf{b}=(b_{1},b_{2},\dots,b_{L})$,
$\mathbf{a}\succ\mathbf{b}$ is defined as $a_{i}>b_{i},\forall i=1,\dots,L$. %
} 
\begin{equation}
\mathbf{p}_{e}^{T}f(\mathbf{p}_{e};\mathbf{h},\mathbf{q})\leq-\kappa\alpha\|\mathbf{p}_{e}\|^{2}\label{eq:contraction-modulas}
\end{equation}
for all $t\geq0$, where $\kappa$ is the step size parameter of the
MWQ iterations in (\ref{eq:alg-pwr-update-0}).
\end{lyxLemQED}
\begin{proof}
Please refer to Appendix \ref{app:prof-lem-conv-speed} for the proof.
\end{proof}

This lemma illustrates that the tracking error term $\mathbf{p}_{e}$
contributes to the negative drift (proportional to $\kappa\alpha$)
in $LV(\mathbf{z})$ in (\ref{eq:LV-VSDS}). The larger the tracking
error $\|\mathbf{p}_{e}\|$ is, the stronger force the MWQ iterations
will drag the system state $\mathbf{p}$ to the optimal $\mathbf{p}^{*}$,
which in turn helps the stabilization of the system. In fact, the
negative drift depends on the MWQ iteration step size $\kappa$, which
controls the \emph{convergence rate}%
\footnote{\label{fn:step-size-latency}Note that the MWQ iteration in (\ref{eq:alg-pwr-update-0})
is expressed in continuous time and a larger $\kappa$ is always desirable
from the perspective of convergence speed. However, in practice, the
MWQ iterations are implemented in discrete time and the corresponding
discrete time step size is given by $\kappa\tau$ where $\tau$ is
the slot duration of iterations. For a given $\tau$, a large $\kappa$
will speed up the iteration but also contributes to a larger \emph{steady
state errors} of $\mathcal{O}(\kappa\tau)$ in the discrete time iterations.%
} of the MWQ iterations under static $\mathbf{h}$ and $\mathbf{q}$. 

From the Lyapunov drift for VSDS in (\ref{eq:LV-VSDS}), the transmission
rate $\hat{\bm{\mu}}(t)$ also contributes to negative drift in (\ref{eq:LV-VSDS}),
which in turns help to stabilize the VSDS. Before we quantify the
negative drift contribution, we first discuss several structural properties
of the transmission rate at the equilibrium. Let $\bm{\mu}^{*}\left(\mathbf{h},\mathbf{q}\right)=\hat{\bm{\mu}}\left(\mathbf{p}^{*}(\mathbf{h},\mathbf{q});\mathbf{h},\mathbf{q}\right)$
be the transmission rate at the equilibrium point $\mathbf{p}^{*}(\mathbf{h},\mathbf{q})$
(optimal transmission rate under the MWQ policy). We have the following
lemmas about the structural property of $\bm{\mu}^{*}(\mathbf{h},\mathbf{q})$
and the actual transmission rate $\hat{\bm{\mu}}(t)$. 
\begin{lyxLemQED}
[Structural Properties of the Transmission Rate at Equilibrium]\label{lem:Solution-property-MWQ}
The transmission rate $\bm{\mu}^{*}(\mathbf{h},\mathbf{q})$ at the
equilibrium $\mathbf{p}^{*}(\mathbf{h},\mathbf{q})$ of the VSDS has
the following properties, 
\begin{equation}
\sum_{l=1}^{L}q_{l}\mu_{l}^{*}(\mathbf{h},\mathbf{q})\geq||\mathbf{q}||\min\left\{ \log\left(\frac{S_{\min}}{V}||\mathbf{q}||\right),L\lambda_{\max}+\log\frac{S_{\min}}{|h_{0}|^{2}}\right\} \label{eq:lem_q-mu-sol-property}
\end{equation}
and 
\begin{equation}
\frac{1}{L}\min\left\{ \log\left(\frac{S_{\min}}{V}||\mathbf{q}||\right),L\lambda_{\max}+\log\frac{S_{\min}}{|h_{0}|^{2}}\right\} \leq\|\bm{\mu}^{*}\left(\mathbf{h},\mathbf{q}\right)\|\leq\log\left(\frac{S_{\max}}{V}\|\mathbf{q}\|\right)\label{eq:lem_mu-property}
\end{equation}
for $\|\mathbf{q}\|S_{\min}>V$.
\end{lyxLemQED}
\begin{proof}
Please refer to Appendix \ref{app:prof-lem-sol-property} for the
proof.
\end{proof}
\begin{lyxLemQED}
[Structural Properties of the Actual Rate $\hat{\bm{\mu}}(t)$]\label{lem:error-trans-rate}
There exists a $\beta>0$ depending on $S_{\min}$ and $S_{\max}$,
such that, for all $t\geq0$, the actual transmission rate at time
$t$, $\hat{\bm{\mu}}(t)=\hat{\bm{\mu}}(\mathbf{p}(t);\mathbf{h}(t),\mathbf{q}(t))$
satisfies 
\begin{equation}
\|\bm{\mu}^{*}(\mathbf{h}(t),\mathbf{q}(t))\|-\log\left(1+\beta\|\mathbf{p}_{e}(t)\|\right)\leq\|\hat{\bm{\mu}}(t)\|\leq\|\bm{\mu}^{*}(\mathbf{h}(t),\mathbf{q}(t))\|+\log\left(1+\beta\|\mathbf{p}_{e}(t)\|\right).\label{eq:mu-tracking-err-rates}
\end{equation}

\end{lyxLemQED}
\begin{proof}
Please refer to Appendix \ref{app:prof-lem-trans-err} for the proof.
\end{proof}
%

%

Lemma \ref{lem:Solution-property-MWQ} shows that the term $\sum q_{l}\mu_{l}^{*}$
grows faster than $\|\mathbf{q}\|$ and $\|\bm{\mu}^{*}(\mathbf{h},\mathbf{q})\|$
is lower bounded with the order $\log(\|\mathbf{q}\|)$. On the other
hand, Lemma \ref{lem:error-trans-rate} illustrates that, although
there is a tracking error $\mathbf{p}_{e}$ in the power allocation,
a minimum transmission rate is still guaranteed and the rate penalty
due to the tracking error $\mathbf{p}_{e}$ is no larger than $\log\left(1+\beta\|\mathbf{p}_{e}\|\right)$.

Based on the properties in Lemma \ref{lem:Solution-property-MWQ}-\ref{lem:error-trans-rate},
we can derive an upper bound of the Lyapunov drift in (\ref{eq:LV-VSDS}).
Let $a_{A}=\|A\|=\max\left\{ a_{l}\right\} $ be the CSI fading rate
parameter, where $A$ is the coefficient matrix of the CSI dynamics
defined in (\ref{sde:CSI-model}). Let $\lambda_{\max}=\|\Lambda\|=\max_{l}\left\{ \lambda_{l}\right\} $
be the maximum arrival rate among all the transmission links. 
\begin{lyxLemQED}
[Lyapunov Drift Property for VSDS]\label{lem:Lyapunov-drift-property}Suppose
there exists constants $\gamma_{q}<\infty$ and $\gamma_{h}<\infty$,
such that $\|\psi_{q}(\mathbf{h}(t),\mathbf{q}(t))\|\leq\gamma_{q}$,
$\|\psi_{h}(\mathbf{h}(t),\mathbf{q}(t))\|\leq\gamma_{h}$, for all
$t\geq0$. In addition, the step size $\kappa$ satisfies $\kappa>\frac{2}{\alpha}\max\left\{ \gamma_{q}^{2},\beta\gamma_{q}\right\} $
under all $\mathbf{h}(t)$. Then the stochastic Lyapunov drift in
(\ref{eq:LV-VSDS}) is bounded by 
\begin{eqnarray}
LV & \leq & -\left(\|\mathbf{p}_{e}\|+\|\mathbf{q}\|+\|\mathbf{h}\|\right)+D(S_{\min},S_{\max})\label{eq:lem-LV-bound}
\end{eqnarray}
where 
\begin{eqnarray*}
D(S_{\min},S_{\max}) & = & L(2a_{A}(1+\gamma_{h}^{2})+\gamma_{q}^{2}+\lambda_{\max})+\frac{1}{8a_{A}[1-\gamma_{h}^{2}a_{A}/(\kappa\alpha)]}\\
 &  & \qquad+\frac{2\gamma_{q}^{2}\lambda_{\max}^{2}}{\kappa\alpha}+\frac{V}{S_{\min}}2^{L\lambda_{\max}-1}+g(S_{\min},S_{\max})+C
\end{eqnarray*}
and $g(S_{\min},S_{\max})$ is a function bounded for all $S_{\min}$
and $S_{\max}$.
\end{lyxLemQED}
\begin{proof}
Please refer to Appendix \ref{app:prof-lem-Lyapunov-drift-property}
for the proof.
\end{proof}

From the above lemma, the Lyapunov drift (\ref{eq:LV-VSDS}) is increasingly
negative for sufficiently large $\|\mathbf{q}\|$ and $\|\mathbf{h}\|$
and this \emph{negative drift }drives the system state back to a trajectory
with bounded norm. This property stabilizes the VSDS. Denote $\overline{\alpha}_{0}=\mathbb{E}\left[\frac{1}{\alpha}\right]$,
$\overline{\gamma}_{0}=\mathbb{E}\left[\frac{1}{8[1-\gamma_{h}^{2}a_{A}/(\kappa\alpha)]}\right]$,
$\overline{\sigma}=\mathbb{E}[\frac{1}{S_{\min}}]$ and $\overline{g}=\mathbb{E}\left[g(S_{\min},S_{\max})\right]$.
Note that as $S_{\min}$ and $S_{\max}$ are the $L$-th order statistics
of stationary processes $|h_{l}|^{2}$, $\overline{\sigma}$ and $\overline{g}$
are bounded above. The stability results of the VSDS can be summarized
as follows.
\begin{lyxThmQED}
[Stability of the VSDS]\label{thm:stability-region-VSDS} The system
state $\mathbf{z}(t)$ of VSDS in (\ref{sde:VSDS}) is stochastically
stable and satisfies 
\begin{equation}
\lim\sup_{t\to\infty}\frac{1}{t}\int_{0}^{t}\mathbb{E}\left[\|\mathbf{z}(\tau)\|\right]d\tau\leq L(2a_{A}(1+\gamma_{h}^{2})+\gamma_{q}^{2}+\lambda_{\max})+\frac{\overline{\gamma}_{0}}{a_{A}}+\frac{\overline{\alpha}_{0}\gamma_{q}^{2}\lambda_{\max}^{2}}{\kappa}+V2^{L\lambda_{\max}-1}\overline{\sigma}+\overline{g}.\label{eq:main-result-queue-bound}
\end{equation}

\end{lyxThmQED}

The above theorem is a direct result of Lemma \ref{lem:Lyapunov-drift-property}
and Theorem \ref{thm:stoch-stability-region-z}. As $\|\mathbf{q}\|\leq\|\mathbf{z}\|$,
we can obtain the average queue bound from the following corollary. 
\begin{lyxCorQED}
[Expected Average Queue Bound under Time-varying CSI and QSI ]\label{cor:Expected-Average-Queue}
The expected average queue bound under MWQ algorithm in time-varying
CSI and QSI is given by 
\begin{equation}
\lim\sup_{t\to\infty}\frac{1}{t}\int_{0}^{t}\mathbb{E}\left[\|\mathbf{q}(\tau)\|\right]d\tau\leq L(2a_{A}(1+\gamma_{h}^{2})+\gamma_{q}^{2}+\lambda_{\max})+\frac{\overline{\gamma}_{0}}{a_{A}}+\frac{\overline{\alpha}_{0}\gamma_{q}^{2}\lambda_{\max}^{2}}{\kappa}+V2^{L\lambda_{\max}-1}\overline{\sigma}+\overline{g}.\label{eq:queue-upper-bound}
\end{equation}

\end{lyxCorQED}

The result shows the upper bound of the average worst case queue (corresponding
to the worst case delay) of the network. The bound depends on several
important parameters, namely the CSI fading rate $a_{A}$, and the
sensitivities of the equilibrium $\mathbf{p}^{*}(\mathbf{h},\mathbf{q})$
w.r.t. $\mathbf{h}$ and $\mathbf{q}$, ($\gamma_{h}$, $\gamma_{q}$). 

\begin{figure}
\begin{centering}
\includegraphics[width=0.9\columnwidth]{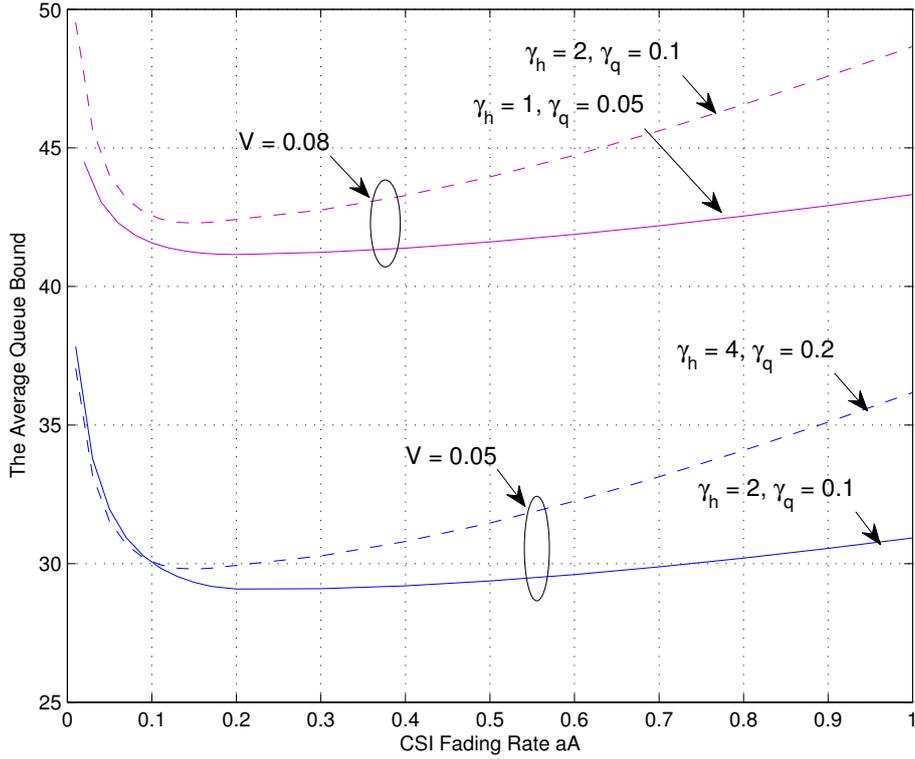}
\par\end{centering}

\caption{\label{fig:queue-bound-J}A numerical illustration of the average
queue bound (for the worst queue) in (\ref{eq:queue-upper-bound})
versus the CSI fading rate $a_{A}$. The numerical result is under
$L=4$ links and maximum arrival rate $\lambda_{\max}=1$, and different
assumptions of sensitivity parameters $\gamma_{h}$ and $\gamma_{q}$.
The numerical queue bound is increasing with the CSI fading rate $a_{A}$,
the parameters $\gamma_{h}$ and $\gamma_{q}$.}
\end{figure}

Fig. \ref{fig:queue-bound-J} gives a numerical illustration of the
theoretical queue bound in (\ref{eq:queue-upper-bound}) under $L=4$
links, maximum arrival rate $\lambda_{\max}=1$,  various CSI fading
rates $a_{A}$, and sensitivity parameters $\gamma_{h}$ and $\gamma_{q}$.
Note that the delay bound increases w.r.t. $\gamma_{h}$ and $\gamma_{q}$.
Note that the delay bound increases w.r.t. $\gamma_{h}$, $\gamma_{q}$
and at both large and small fading speed ($a_{A}$). For large $a_{A}$,
there is the penalty of the increased tracking error due to time varying
CSI. For small $a_{l}$, the delay increases because the CSI may be
stuck at a poor state for quite a long time. 

\section{Adaptive Compensation for the MWQ Algorithm in Time-varying Arrivals
and Channels}

Based on the stochastic dynamics modeled by the VSDS, we consider
modifying the gradient MWQ iterations in (\ref{eq:alg-pwr-update-0})
to reduce the penalty induced by time varying CSI and QSI. Specifically,
we introduce a \emph{compensation term} to improve the stochastic
dynamics of the VSDS in (\ref{sde:VSDS}). This corresponds to a\emph{
compensation} term in the MWQ algorithm to offset the effect from
the time-varying CSI and QSI. The overall compensated MWQ algorithm
is shown to have a better convergence robustness w.r.t. time varying
CSI and QSI both analytically and numerically.

\subsection{A Proposed Algorithm with Compensation Term}

We have shown in Section \ref{sub:VSDS-Virtual-Stochastic-Dynamic}
that the dynamics of tracking error $\mathbf{p}_{e}(t)$ can be modeled
by a stochastic error dynamic system in (\ref{sde:power-error-dynamic}),
which consists of a drift term $f_{e}(\centerdot)$ and diffusion
terms $b_{e}(\centerdot)$ and $c_{e}(\centerdot;d\mathbf{W})$. Without
the diffusion terms, the SEDS eventually converges to the origin,
as $f_{e}(\centerdot)$ contributes a negative drift to the infinitesimal
estimator $LV(\mathbf{p}_{e})$, where we define the Lyapunov function
as $V(\mathbf{p}_{e})=\mathbf{p}_{e}^{T}\mathbf{p}_{e}$. However,
with the presence of the diffusion terms, the system $d\mathbf{p}_{e}=f_{e}(\mathbf{p}_{e};\mathbf{h},\mathbf{q})dt$
is disturbed from the equilibrium at $\mathbf{p}_{e}=\mathbf{0}$
and the state $\mathbf{p}_{e}$ is driven away from the origin. The
magnitude of $b_{e}(\centerdot)$ and $c_{e}(\centerdot)$ reflect
the chance and intensity that the state $\mathbf{p}_{e}(t)$ is being
disturbed. Based on this observation, one way to stabilize $\mathbf{p}_{e}(t)$
is to offset the diffusion terms $b_{e}(\centerdot)$ and $c_{e}(\centerdot)$
in the SEDS dynamics in (\ref{sde:power-error-dynamic}). Equivalently,
this corresponds to modifying the MWQ algorithm iterations in (\ref{eq:alg-pwr-update-0})
to compensate for the effects of time varying CSI and QSI. From the
error tracking vector\textbf{ $d\mathbf{p}_{e}$} in (\ref{eq:dpe}),
we would like to compensate the movement of the optimal target $d\mathbf{p}^{*}(t)$
so that the resulting SEDS becomes $d\mathbf{p}_{e}=\kappa\left[\nabla\mathcal{L}\left(\mathbf{p}_{e}+\mathbf{p}^{*};\mathbf{h},\mathbf{q}\right)\right]_{\mathbf{p}_{e}}^{\mathcal{P}}dt$.
In this ideal case, the $\mathbf{p}_{e}$ will converge to $\mathbf{0}$.
However, the challenge is that we do not have an exact expression
for $d\mathbf{p}^{*}(t)$ during the iteration because we do not have
closed form expression of the equilibrium $\mathbf{p}^{*}(t)$. We
shall propose an indirect method of estimating the compensation term. 

Since the MWQ problem in (\ref{eq:prob-power-allocation}) is convex,
$\mathbf{p}^{*}$ is the optimum if and only if there exists $\bm{\lambda}^{*}\succeq\mathbf{0}$,
such that 
\begin{eqnarray}
\nabla\mathcal{L}(\mathbf{p}^{*};\mathbf{h},\mathbf{q})+\bm{\lambda}^{*} & = & \mathbf{0}\label{eq:KKT-L}\\
\lambda_{l}^{*}p_{l}^{*} & = & 0\quad\forall l=1,\dots,L.\label{eq:KKT-r}
\end{eqnarray}
We denote the above system of equations (KKT conditions) as $\Phi(\mathbf{x}^{*};\mathbf{h},\mathbf{q})=\mathbf{0}$,
where $\mathbf{x}^{*}=(\mathbf{p}^{*},\bm{\lambda}^{*})$. Note that
$\mathbf{x}^{*}$ is unique for a convex problem. Using implicit function
theorem and assuming $\frac{\partial\Phi}{\partial\mathbf{x}^{*}}$
is non-singular, we have 
\begin{equation}
d\mathbf{x}^{*}=\left[\begin{array}{c}
d\mathbf{p}^{*}\\
d\bm{\lambda}^{*}
\end{array}\right]=-\left(\frac{\partial\Phi}{\partial\mathbf{x}^{*}}\right)^{-1}\frac{\partial\Phi}{\partial\mathbf{q}}d\mathbf{q}-2\mbox{Re}\left[\left(\frac{\partial\Phi}{\partial\mathbf{x}^{*}}\right)^{-1}\frac{\partial\Phi}{\partial\mathbf{h}}d\mathbf{h}\right].\label{eq:com-dx}
\end{equation}
As a result, we obtain $d\mathbf{p}^{*}=\hat{\varphi}_{q}(\mathbf{p}^{*},\bm{\lambda}(\mathbf{p});\mathbf{h},\mathbf{q})d\mathbf{q}+\mbox{Re}\left[\hat{\varphi}_{h}(\mathbf{p}^{*},\bm{\lambda}(\mathbf{p});\mathbf{h},\mathbf{q})d\mathbf{h}\right]$,
where the vector-valued functions $\hat{\varphi}_{q}(\mathbf{p}^{*};\centerdot)$
and $\hat{\varphi}_{h}(\mathbf{p}^{*};\centerdot)$ are the rows for
primal variable $d\mathbf{p}^{*}$ from $-\left(\frac{\partial\Phi}{\partial\mathbf{x}^{*}}\right)^{-1}\frac{\partial\Phi}{\partial\mathbf{q}}$
and $-2\left(\frac{\partial\Phi}{\partial\mathbf{x}^{*}}\right)^{-1}\frac{\partial\Phi}{\partial\mathbf{h}}$
in (\ref{eq:com-dx}), respectively. Thus the MWQ iterations with
compensation is given by 
\begin{equation}
\dot{\mathbf{p}}=\left[\kappa\nabla\mathcal{L}(\mathbf{p};\mathbf{h}(t),\mathbf{q}(t))-\hat{\varphi}_{q}(\mathbf{p},\bm{\lambda}(\mathbf{p});\mathbf{h}(t),\mathbf{q}(t))d\mathbf{q}-\mbox{Re}\left[\hat{\varphi}_{h}(\mathbf{p},\bm{\lambda}(\mathbf{p});\mathbf{h}(t),\mathbf{q}(t))d\mathbf{h}\right]\right]_{\mathbf{p}}^{\mathcal{P}}\label{eq:compentation-alg}
\end{equation}
where $\hat{\varphi}_{q}(\centerdot)d\mathbf{q}$ and $\mbox{Re}\left[\hat{\varphi}_{h}(\centerdot)d\mathbf{h}\right]$
are compensation terms. Here, we use the current algorithm state $\mathbf{p}(t)$
as an approximation of the target equilibrium $\mathbf{p}^{*}(t)$
and $\bm{\lambda}$ is computed via the KKT conditions in (\ref{eq:KKT-L})-(\ref{eq:KKT-r}).
The compensation term can be interpreted as an estimation on how the
target equilibrium $\mathbf{p}^{*}$ is moving according to the time-varying
CSI and QSI $(d\mathbf{h},d\mathbf{q})$. When $\mathbf{p}$ is close
to $\mathbf{p}^{*}$ (i.e., $\mathbf{p}_{e}$ is small), the estimation
$\hat{\varphi}(\mathbf{p},\bm{\lambda}(\mathbf{p});d\mathbf{h},d\mathbf{q})$
on $d\mathbf{p}^{*}$ is accurate. Thus the compensation term helps
further reduce the tracking error and the algorithm would eventually
converge to the equilibrium $\mathbf{p}^{*}$. We shall investigate
the convergence behavior of the compensation algorithm in the following
subsection.

\subsection{Performance Analysis for the Compensation Algorithm}

Suppose the functions $\hat{\varphi}_{q}(\mathbf{p};\centerdot)$
and $\hat{\varphi}_{h}(\mathbf{p};\centerdot)$ are Lipschitz continuous,
i.e., there exists positive constants $L_{q}$, $L_{h}<\infty$, such
that $\|\hat{\varphi}_{q}(\mathbf{p};\centerdot)-\hat{\varphi}_{q}(\mathbf{p}^{*};\centerdot)\|\leq L_{q}\|\mathbf{p}-\mathbf{p}^{*}\|$
and $\|\hat{\varphi}_{h}(\mathbf{p};\centerdot)-\hat{\varphi}_{h}(\mathbf{p}^{*};\centerdot)\|\leq L_{h}\|\mathbf{p}-\mathbf{p}^{*}\|$,
for all $\mathbf{p}\in\mathbb{R}_{+}^{L}$. Let $\mu_{\max}$ be the
maximum transmission rate that the system can support and $\alpha>0$
be defined in (\ref{eq:contraction-modulas}) uniformly for all CSI
realization $\mathbf{h}$. The following theorem provides a sufficient
condition to the convergence of the compensation algorithm.

\begin{lyxThmQED}
[Convergence of the Compensation Algorithm]\label{thm:convergence-compensation}
Provided that the step size parameter $\kappa$ satisfies, 
\begin{eqnarray*}
\kappa & > & \frac{1}{\alpha}\left[\left(\mu_{\max}+\lambda_{\max}L\right)L_{q}+\frac{1}{2}L_{q}^{2}+\frac{1}{2}a_{A}L_{h}^{2}\right]
\end{eqnarray*}
for all $t\geq0$. Then the MWQ iterations with compensation in (\ref{eq:compentation-alg})
asymptotically tracks the moving equilibrium point $\mathbf{p}^{*}(t)$
with no errors, i.e., $\forall\epsilon>0$, 
\[
\lim_{t\to0}\mbox{Pr}\left[\|\mathbf{p}(t)-\mathbf{p}^{*}(t)\|<\epsilon\right]=1.
\]

\end{lyxThmQED}
\begin{proof}
Please refer to Appendix \ref{app:Proof-thm-convergence-compensation}
for the proof.
\end{proof}

Theorem \ref{thm:convergence-compensation} shows that when a large
enough step size $\kappa$ is available, the compensation algorithm
can converge to the equilibrium point $\mathbf{p}^{*}(t)$, and there
is no performance penalty due to the time-varying CSI and QSI. The
convergence is affected by the parameters $L$, $a_{A}$, $L_{h}$
and $L_{q}$, where $L$ is number of transmission links in the network
(the system dimension), $a_{A}$ is the CSI variation rate of the
whole network, and $L_{h}$ and $L_{q}$ represent the sensitivity
of the equilibrium point $\mathbf{p}^{*}(t)$ w.r.t. the time-varying
CSI and QSI. On the other hand, for conventional gradient iteration
in (\ref{eq:alg-pwr-update-0}), the algorithm cannot have $\mathbf{p}_{e}\to\mathbf{0}$
no matter how large the iteration step size $\kappa$ is used. This
is due to the fact that the target equilibrium $\mathbf{p}^{*}(t)$
is moving due to the time-varying CSI and QSI. 
\begin{remrk}
[Interpretation of the results] In practice, we would like to implement
the modified MWQ iteration in (\ref{eq:compentation-alg}) on discrete
time. The iterations of (\ref{eq:compentation-alg}) can be written
as 
\[
\mathbf{p}(t+\tau)=\left\{ \mathbf{p}(t)+\kappa\tau\nabla\mathcal{L}(\mathbf{p}(t);\mathbf{h}(t),\mathbf{q}(t))-\hat{\varphi}_{q}(\centerdot)\triangle\mathbf{q}(t)-\mbox{Re}\left[\hat{\varphi}_{h}(\centerdot)\triangle\mathbf{h}(t)\right]\right\} _{\mathbf{p}}^{\mathcal{P}}
\]
in discrete time where $\triangle x(t)=x(t+\tau)-x(t)$ and $\tau$
is the time step. In this case, the overall error between $\mathbf{p}(t)$
and $\mathbf{p}^{*}(t)$ is contributed by (a) algorithm convergence
error and (b) steady state error. While Theorem \ref{thm:convergence-compensation}
suggests that a large step size $\kappa$ is always desirable from
the algorithm convergence error perspective, the above analysis did
not consider the \emph{steady state error} (due to constant step size)
$o(\kappa\tau)$ associated with discrete-time implementation. The
overall impacts of steady state errors and tracking errors will be
demonstrated in the numerical results section. 

\end{remrk}

\section{Numerical Results and Discussions}

In this section, we shall simulate the tracking performance of the
conventional MWQ iteration and the proposed compensated MWQ iteration
in time-varying channels. We also demonstrate the delay performance
for the two MWQ iterations under various CSI fading rates. We consider
a wireless ad-hoc network with $5$ nodes and $6$ links as depicted
in Fig. \ref{fig:Network-topology}. The $l$-th link transmits the
$l$-th data flow. Transmission flows towards a same destination share
the same frequency band and SIC is implemented at each receiving node
to handle the inter-flow interference. The CSI $h_{l}$ for each link
is modeled by a unit variance Markov process described by the SDE
in (\ref{sde:CSI-model}). Data arrivals are modeled by continuous
time Poisson processes with rate $\lambda=20$ packets/second. All
the algorithms are implemented in discrete-time iterations with  simulation
time step $1$ ms and the queueing system was run over a time duration
$T=100$ min. The delay performance of the conventional MWQ iterations
in (\ref{eq:alg-pwr-update-0}) and the modified MWQ iterations with
compensation in (\ref{eq:compentation-alg}) are compared against
the following reference baselines.
\begin{itemize}
\item \textbf{Baseline 1 - Constant Power Allocation:} At each time slot,
fixed power $P$ is allocated to each link and the transmission rate
is computed by (\ref{eq:mu-hat-1})-(\ref{eq:mu-hat-k}).
\item \textbf{Baseline 2 - Throughput Optimal Power Allocation:} The throughput
optimal power control is computed by solving the MWQ problem in (\ref{eq:prob-weighted-queue})-(\ref{eq:prob-const-capacity-region})
to obtain the target equilibrium $\mathbf{p}^{*}(\mathbf{h},\mathbf{q})$
at each time slot $t$. 
\end{itemize}

\subsection{Power Tracking Performance of the MWQ Iterations}

\begin{figure}
\begin{centering}
\includegraphics[width=0.9\columnwidth]{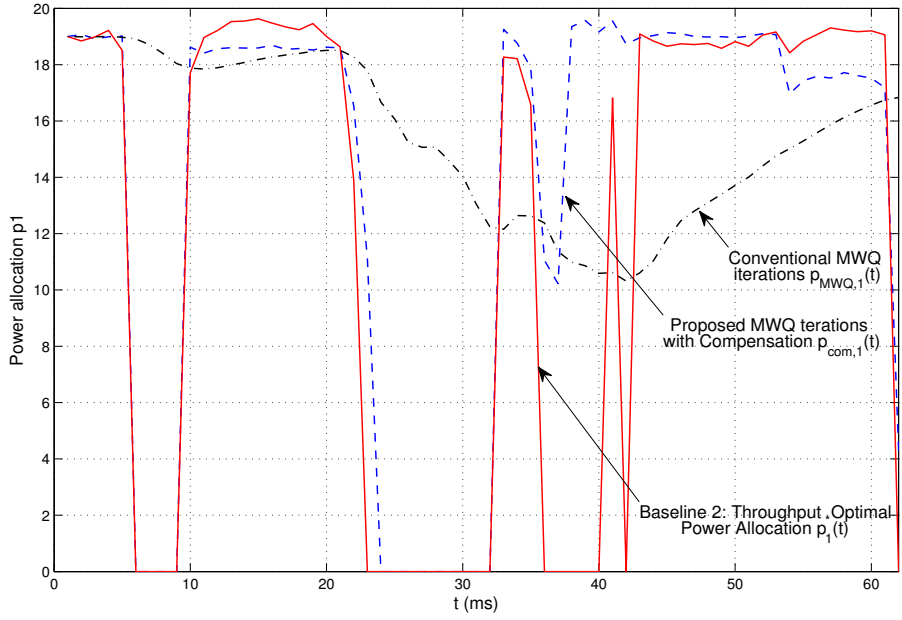}
\par\end{centering}

\caption{\label{fig:power-trajectory}The power control algorithm trajectory
$\mathbf{p}(t)$ versus time at a CSI fading rate $a_{A}$ = 200 and
packet arrival rate $\lambda=20$ packets/second. The algorithms update
on every $\tau=1$ ms time slot with step size $0.5$ (corresponding
to $\kappa=500\mbox{ sec}^{-1}$). The average delay is measured to
be $\overline{T}_{l}\approx500$ ms. As illustrated, the target equilibrium
$p_{1}^{*}(t)$ changes significantly over time due to the time varying
CSI. The conventional MWQ iterations $p_{MWQ,1}(t)$ fail to track
the moving target $p_{1}^{*}(t)$ accurately but the trajectory of
the compensated MWQ iterations $p_{com,1}(t)$ can track the moving
target quite well. }
\end{figure}

Fig. \ref{fig:power-trajectory} captures the power control algorithm
trajectory $\mathbf{p}(t)$ versus time at a CSI fading rate of $a_{A}$
= 200. The algorithms update on every $\tau=1$ ms time slot and the
step size is chosen to be $0.5$ (corresponding to $\kappa=500\mbox{ sec}^{-1}$
for continuous-time trajectory). Throughout the simulation, the average
delay is measured as $\overline{T}_{l}\approx500$ ms. As illustrated,
the target equilibrium $p_{1}^{*}(t)$ changes significantly over
time due to the time varying CSI. The conventional MWQ iterations
$p_{MWQ,1}(t)$ fail to track the moving target $p_{1}^{*}(t)$ accurately
but the trajectory of the compensated MWQ iterations $p_{com,1}(t)$
can track the moving target quite well. 

\begin{figure}
\begin{centering}
\includegraphics[width=0.9\columnwidth]{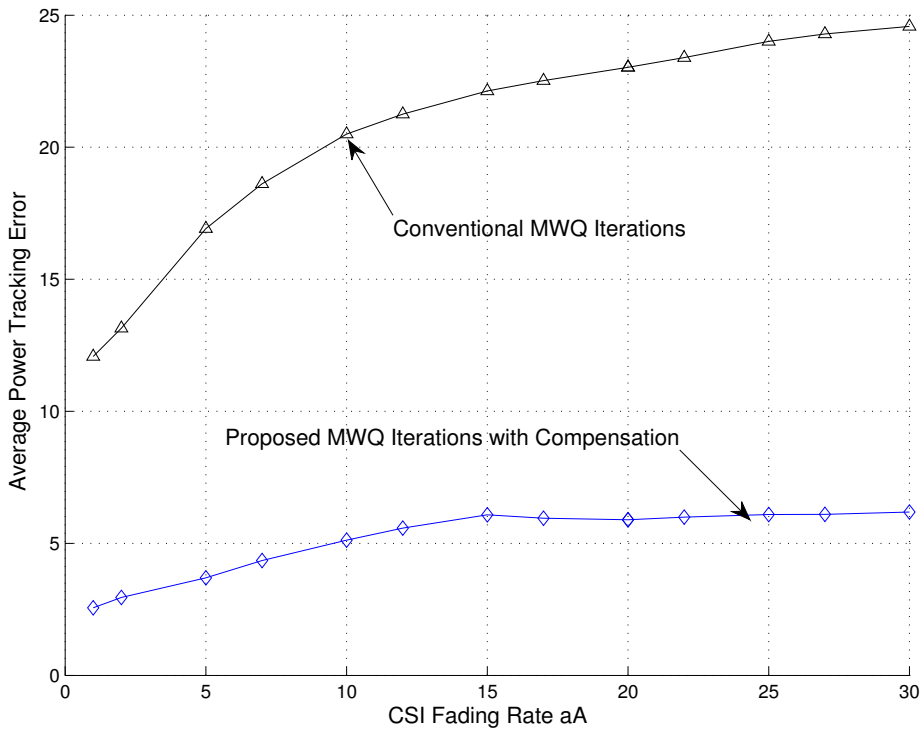}
\par\end{centering}

\caption{\label{fig:power-tracking-error}The average tracking error of the
power trajectory $\mathbf{p}(t)$ versus the fading rate $a_{A}$
under packet arrival rate $\lambda=20$ packets/second. The algorithms
update on every $\tau=1$ ms time slot with step size $0.5$ (corresponding
to $\kappa=500\mbox{ sec}^{-1}$). The average tracking error of conventional
MWQ iterations increases with the fading rate $a_{A}$. On the other
hand, the tracking error of the modified MWQ iterations (with compensations)
is much smaller than that of the conventional MWQ iterations. Note
that the error consists of contributions from both the algorithm convergence
error and steady state error due to constant step size (in discrete
time). From Theorem \ref{thm:convergence-compensation}, the algorithm
convergence error of the modified MWQ converges to zero but there
is still residual steady state error. }
\end{figure}

Fig. \ref{fig:power-tracking-error} illustrates the average tracking
error of the power trajectory $\mathbf{p}(t)$ versus the fading rate
$a_{A}$. The average tracking error of the power trajectory is defined
as $e=\frac{1}{T}\int_{0}^{T}\|\mathbf{p}(t)-\mathbf{p}^{*}(t)\|dt$.
It is shown that the average tracking error of conventional MWQ iterations
increases with the fading rate $a_{A}$. On the other hand, the tracking
error of the modified MWQ iterations (with compensations) is much
smaller%
\footnote{Note that the tracking error shown is the overall error obtained using
discrete-time iterations, which include the errors due to algorithm
convergence and \emph{steady state errors} (due to constant discrete
time step size). From Theorem \ref{thm:convergence-compensation},
the algorithm convergence error tends to zero for the modified MWQ
but there is a steady state error in Fig. \ref{fig:power-tracking-error}
due to the constant step size in discrete time implementation. %
} than that of the conventional MWQ iterations. 

\subsection{Power-Delay Tradeoff Performance}

\begin{figure}
\begin{centering}
\includegraphics[width=0.9\columnwidth]{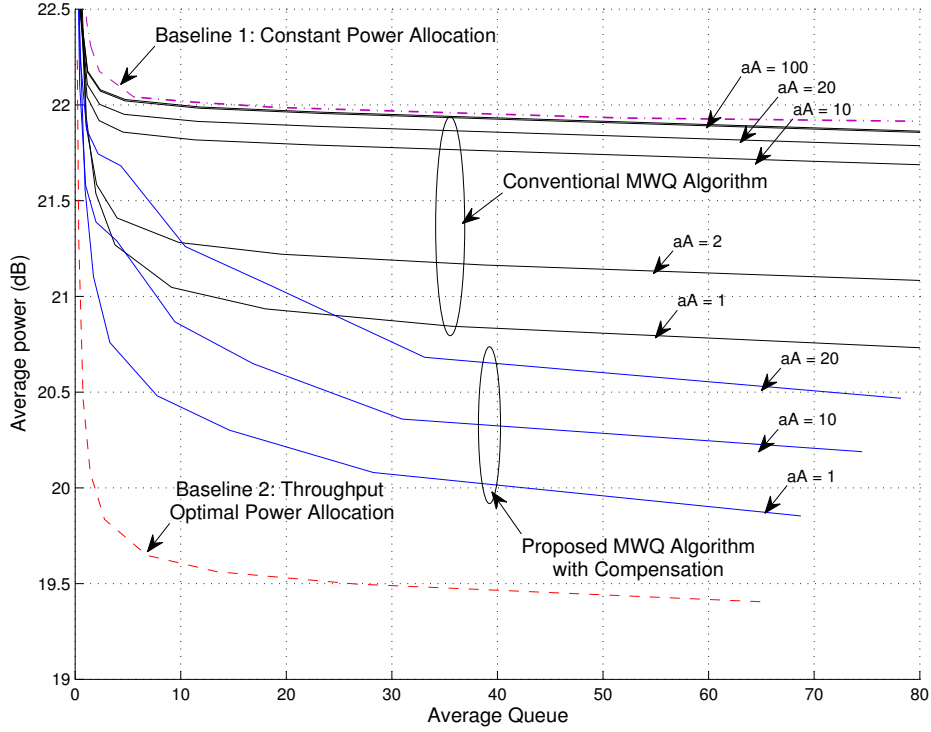}
\par\end{centering}

\caption{\label{fig:pwr-delay-tradeoff}The per-node average power versus the
average delay at different fading rate. Observed that to maintain
the same average delay of 2 seconds, the conventional MWQ iterations
require 2.3 dB more power than the throughput optimal scheme. On the
other hand, the proposed modified MWQ algorithm with compensation
suffers from a very small power penalty (< 1dB) compared with baseline
2 (the throughput optimal scheme). Furthermore, as the CSI fading
rate $a_{A}$ increases, the conventional MWQ iterations eventually
require as much power as baseline 1 (constant power allocation) does,
while the proposed modified MWQ algorithm with compensation still
has a reasonable power gain compared to baseline 1.}
\end{figure}

Fig. \ref{fig:pwr-delay-tradeoff} illustrates the per-node average
power versus the average delay at different fading rates. Note that
along each curve, we have different values of $V$, which acts as
a tradeoff parameter for power-delay tradeoff. Small $V$ corresponds
to small delay and vice versa. Observed that to maintain the same
average delay of 2 seconds, the conventional MWQ iterations require
2.3 dB more power than the throughput optimal scheme. On the other
hand, the proposed modified MWQ algorithm with compensation suffers
from a very small power penalty (< 1dB) compared with baseline 2 (the
throughput optimal scheme). Furthermore, as the CSI fading rate $a_{A}$
increases, the conventional MWQ iterations eventually require as much
power as baseline 1 (constant power allocation) does, while the proposed
modified MWQ algorithm with compensation still has a reasonable power
gain compared to baseline 1.

\section{Conclusions}

In this paper, we have analyzed the convergence behavior and the queue
delay performance of the conventional MWQ iterations in a wireless
adhoc network, in which the CSI and the QSI are changing in a similar
timescale as the algorithm iterations. We first show that the algorithm
convergence can be captured by studying the stochastic stability of
an equivalent \emph{virtual stochastic dynamic system} (VSDS). By
extending the Foster-Lyapunov criteria, we established the technical
conditions for queue stability and derived the associated queue bounds.
Based on these analyses, we have proposed a novel adaptive MWQ algorithm
with a predictive compensation to counteract the effects of the time
varying CSI and QSI. We have demonstrated that with some mild conditions,
the modified MWQ iterations (with compensation) can converge to the
moving target power $\mathbf{p}^{*}(t)$ despite the time varying
CSI and QSI. Finally, simulation results demonstrated the performance
gain of the proposed algorithm in both the network delay performance
and the tracking error of the power trajectory. 

\appendices

\section{Connections between the Optimization Algorithms and the VSDS\label{app:thm1-connection}}

In this section, we give a brief introduction to the Lyapunov method
for algorithm convergence analysis, which motivates us to connect
the algorithm trajectory to the VSDS. 

We focus on gradient-based methods that are widely used for computing
the optimal resource allocations in wireless communication networks
and are well-suited for implementations across a distributed network.
The gradient method searches the optimum point $x^{*}$ of the objective
function $\mathcal{L}(x)$ following 
\[
\dot{x}=\frac{dx}{dt}=\kappa\left[\frac{\partial\mathcal{L}}{\partial x}\right]^{T}.
\]
Here we study the convergence behavior by constructing the tracking
\emph{error dynamics} of the algorithm trajectory. Define the tracking
error $x_{e}=x-x^{*}$ and substitute it into the above dynamics,
we obtain 
\begin{eqnarray}
\dot{x}_{e} & = & \kappa\left[\frac{\partial\mathcal{L}(x_{e}+x^{*})}{\partial x_{e}}\right]^{T}\triangleq f(x_{e}).\label{eq:app-virtual-error-dynamic-1}
\end{eqnarray}
Hence the convergence analysis is transferred to stability analysis
\cite{Khalil1996} of the virtual error dynamic system (\ref{eq:app-virtual-error-dynamic-1})
at the origin $x_{e}=0$. 

A classic method to study the stability of a dynamic system is via
the Lyapunov theory \cite{Khalil1996}. We first construct a Lyapunov
function which has the following properties,
\[
V(x_{e})\to\infty,\;\mbox{as}\;\|x_{e}\|\to\infty,\quad\mbox{and }V(x_{e})\to0,\;\mbox{as}\;\|x_{e}\|\to0.
\]
The Lyapunov theory says, if $\dot{V}(x_{e})<0$ for all $x_{e}\in\mathbb{R}^{n}\backslash\{0\}$,
then the dynamic system $\dot{x}_{e}=f(x_{e})$ is asymptotically
stable at the origin $x_{e}=0$ \cite{Khalil1996}. 

Note that, the objective function $\mathcal{L}(x;h(t),q(t))$ we focus
on in this paper has stochastic time-varying parameters $h(t)$ and
$q(t)$, which may evolve in a similar timescale to the algorithm
trajectory. We tackle this problem by constructing the VSDS from the
algorithm dynamics, and extending the Foster-Lyapunov criteria (in
Theorem \ref{thm:stoch-stability-region-z}). We show the connection
between the algorithm trajectory and the VSDS in the following.

\emph{Proof of Theorem \ref{Thm: Connections-virtual-algorithm}}:
Note that the VSDS in (\ref{sde:VSDS}) consists of three components,
\textbf{$\mathbf{p}_{e}$}, $\mathbf{h}$ and $\mathbf{q}$, where
the dynamics of $\mathbf{h}(t)$\textbf{ }and $\mathbf{q}(t)$ are
just the same as (\ref{SDE:system-dynamic-0-h}) and (\ref{SDE:system-dynamic-0-q}).
We only need to show that the dynamics of $\mathbf{p}_{e}(t)$ in
the VSDS in (\ref{sde:VSDS}) implies the MWQ power control algorithm
dynamics of $\mathbf{p}(t)$ in (\ref{SDE:system-dynamic-0-p}). Equivalently,
we need to show 
\[
\mathbf{p}(t)=\mathbf{p}_{e}(0)+\int_{0}^{t}d\mathbf{p}_{e}(\tau)+\mathbf{p}^{*}(t)
\]
to be the solution of (\ref{SDE:system-dynamic-0-p}). On the other
hand, by the definition of tracking error (Definition \ref{def:Tracking-Error}),
$\mathbf{p}_{e}(0)+\int_{0}^{t}d\mathbf{p}_{e}(\tau)+\mathbf{p}^{*}(t)=\mathbf{p}_{e}(t)+\mathbf{p}^{*}(t)=\mathbf{p}(t)$.
Therefore, by substituting $\mathbf{p}_{e}(t)+\mathbf{p}^{*}(t)$
with $\mathbf{p}(t)$ in the VSDS in (\ref{sde:VSDS}), we see that
the trajectory $\mathbf{q}(t)$ in the VSDS is just the same as that
in (\ref{SDE:system-dynamic-0-q}).

To prove the second part of the theorem, we consider that there is
no disturbance applied to the SEDS in (\ref{sde:power-error-dynamic})
by considering $ $$d\mathbf{N}\equiv\mathbf{0}$ and $d\mathbf{W}\equiv\mathbf{0}$.
Equivalently, we take $d\mathbf{h}=\mathbf{0}$ and $d\mathbf{q}=\mathbf{0}$
in (\ref{eq:dpe}). The SDE of $\mathbf{p}_{e}(t)$ in (\ref{sde:power-error-dynamic})
reduces to 
\begin{eqnarray}
d\mathbf{p}_{e} & = & \kappa\left[\nabla\mathcal{L}\left(\mathbf{p}_{e}+\mathbf{p}^{*};\mathbf{h},\mathbf{q}\right)\right]_{\mathbf{p}_{e}+\mathbf{p}^{*}}^{+}dt.\label{eq:sys-pwr-error-static}
\end{eqnarray}
By the definition of equilibrium point (Definition \ref{def:Equilibrium-Point}),
$\mathbf{p}_{e}\to\mathbf{0}$ corresponds to $\nabla\mathcal{L}\to\mathbf{0}$.
Hence the origin is an equilibrium to the SDE in (\ref{sde:power-error-dynamic}).
On the other hand, if the origin is an equilibrium to (\ref{eq:sys-pwr-error-static}),
$\mathbf{p}^{*}=\psi(\mathbf{h},\mathbf{q})$ must be the equilibrium
to the dynamics of the power control algorithm in (\ref{SDE:system-dynamic-0-p}).
Hence, we complete the proof.

\section{Proof of Theorem \ref{thm:stoch-stability-region-z} \label{app:lem-stochastic-stability-region-z}}
\begin{proof}
Define a sequence of stopping time $t_{n}=\inf\left\{ t\geq0:\mathbf{z}(t)\geq n\right\} $.
By Dynkin's formula \cite{Mao:SDE1997}, 
\[
0\leq V(\mathbf{z}(t_{n}))\leq V(\mathbf{z}(0))+\mathbb{E}\left[\int_{0}^{t_{n}}\left(-a\|\mathbf{z}(\tau)\|+g(\mathbf{s}(\tau))\right)d\tau\right].
\]
Hence we have 
\[
\mathbb{E}\left[\int_{0}^{t_{n}}a\|\mathbf{z}(\tau)\|d\tau\right]\leq V(\mathbf{z}(0))+\mathbb{E}\left[\int_{0}^{t_{n}}g(\mathbf{s}(\tau))d\tau\right]
\]
Exchanging the order of integration and expectation, we have 
\[
\frac{1}{t_{n}}\int_{0}^{t_{n}}\mathbb{E}\|\mathbf{z}(\tau)\|d\tau\leq\frac{1}{t_{n}}\frac{V(\mathbf{z}(0))}{a}+\frac{1}{t_{n}}\int_{0}^{t_{n}}\mathbb{E}\left[g(\mathbf{s}(\tau))\right]d\tau
\]
Taking limit on both sides, we obtain 
\[
\lim\sup_{n\to\infty}\frac{1}{t_{n}}\int_{0}^{t_{n}}\mathbb{E}\|\mathbf{z}(\tau)\|d\tau\leq\lim\sup_{n\to\infty}\left(\frac{1}{t_{n}}\frac{V(\mathbf{z}(0)}{a}+\frac{1}{t_{n}}\int_{0}^{t_{n}}\frac{1}{a}\mathbb{E}\left[g(\mathbf{s}(\tau))\right]\right)\leq\frac{d}{a}.
\]
Notice that $t_{n}\to\infty$ as $n\to\infty,$ and $V(\mathbf{z}(0))$
is bounded. Thus the result holds.
\end{proof}

\section{Proof of Lemma \ref{lem:conv-speed-gradient-alg}\label{app:prof-lem-conv-speed}}
\begin{proof}
According to Lemma \ref{lem:Optimal-rate-allocation}, the optimization
problem (\ref{eq:prob-power-allocation}) can be written as 
\begin{eqnarray}
 &  & \max_{\mathbf{p}\in\mathcal{P}}\quad q_{\pi(1)}\log\left(1+|h_{\pi(1)}|^{2}p_{\pi(1)}\right)\label{eq:log-objective-function}\\
 &  & \qquad\qquad+q_{\pi(2)}\left[\log\left(1+\left|h_{\pi(1)}\right|^{2}p_{\pi(1)}+\left|h_{\pi(2)}\right|^{2}p_{\pi(2)}\right)-\log\left(1+|h_{\pi(1)}|^{2}p_{\pi(1)}\right)\right]+\dots\nonumber \\
 &  & \qquad\qquad+q_{\pi(L)}\left[\log\left(1+\sum_{i=1}^{L}\left|h_{\pi(i)}\right|^{2}p_{\pi(i)}\right)-\log\left(1+\sum_{i=1}^{L-1}\left|h_{\pi(i)}\right|^{2}p_{\pi(i)}\right)\right]-\sum_{i=1}^{L}Vp_{\pi(i)}\nonumber 
\end{eqnarray}
for a certain permutation $\bm{\pi}$, where $q_{\pi(k-1)}\geq q_{\pi(k)}$,
for $k=2,\dots,L$. As the objective function $\mathcal{L}(\mathbf{p};\mathbf{h},\mathbf{q})$
is a combination of logarithmic functions, it can be verified that
$\mathcal{L}(\mathbf{p};\mathbf{h},\mathbf{q})$ is strictly concave
in $\mathbf{p}$ and $\nabla^{2}\mathcal{L}(\mathbf{p};\mathbf{h},\mathbf{q})<0$.
In addition, as the domain $\mathcal{P}$ is compact and under the
condition that $q_{\pi(i)}\geq1$, there exists a positive constant
$\alpha(S_{\min},S_{\max})>0$ depending only on the channel gain
parameters $S_{\min}(t)$ and $S_{\max}(t)$, such that the Hessian
of $\mathcal{L}(\mathbf{p};\mathbf{h},\mathbf{q})$ satisfies $\nabla^{2}\mathcal{L}\preceq-\alpha\mathbf{I}$
for all $\mathbf{p}\in\mathcal{P}$.

Based on this observation, we obtain, 
\begin{eqnarray}
\mathbf{p}_{e}^{T}f(\mathbf{p}_{e};\mathbf{h},\mathbf{q}) & = & \mathbf{p}_{e}^{T}f(\mathbf{0}_{+};\mathbf{h},\mathbf{q})+\mathbf{p}_{e}^{T}\int_{0}^{1}\nabla f(\xi\mathbf{p}_{e};\mathbf{h},\mathbf{q})d\xi\mathbf{p}_{e}\label{eq:equality-1}\\
 & = & \mathbf{p}_{e}^{T}f(\mathbf{0}_{+};\mathbf{h},\mathbf{q})+\mathbf{p}_{e}^{T}\int_{0}^{1}\kappa\nabla^{2}\mathcal{L}(\xi\mathbf{p}_{e}+\mathbf{p}^{*}(\mathbf{h},\mathbf{q});\mathbf{h},\mathbf{q})d\xi\mathbf{p}_{e}\label{eq:equality-2}\\
 & \leq & -\int_{0}^{1}\alpha\kappa\|\mathbf{p}_{e}\|^{2}d\xi\label{eq:inequality-3}\\
 & = & -\alpha\kappa\|\mathbf{p}_{e}\|^{2}\nonumber 
\end{eqnarray}
where $\mathbf{p}_{e}^{T}f(\mathbf{0}_{+};\mathbf{h},\mathbf{q})=(\mathbf{p}-\mathbf{p^{*})}^{T}\nabla\mathcal{L}(\mathbf{p}^{*}(\mathbf{h},\mathbf{q});\mathbf{h},\mathbf{q})\leq0$
is the optimality condition for $\mathbf{p}^{*}(\mathbf{h},\mathbf{q})$
in the optimization problem (\ref{eq:prob-power-allocation}). The
equality (\ref{eq:equality-1})is from Taylor expansion of the gradient
iteration function $f(\centerdot)$, the second equality (\ref{eq:equality-2})
is from the fact that $\nabla f=\nabla^{2}\mathcal{L}$, since $f=\nabla\mathcal{L}$,
and the inequality (\ref{eq:inequality-3}) is from $\nabla^{2}\mathcal{L}\preceq-\alpha\mathbf{I}$
derived above. Hence we proved the result.
\end{proof}

\section{Proof of Lemma \ref{lem:Solution-property-MWQ} \label{app:prof-lem-sol-property}}
\begin{proof}
We first consider a time division MWQ policy. At each time slot, only
the link is selected for transmission and the policy is given in the
following \cite{Neely06-Energy}.

1) Find a link $\hat{l}$ such that 
\begin{equation}
\hat{l}=\arg\max_{l=1,\dots,L}\left\{ q_{l}\log(1+\left|h_{l}\right|^{2}p_{l})-Vp_{l}\right\} \label{eq:app-tdm-link}
\end{equation}

2) \emph{Power allocation:} the power $\mathbf{p}$ is allocated according
to 
\begin{equation}
\widetilde{p}_{l}=\begin{cases}
\left(\frac{q_{l}}{V}-\frac{1}{\left|h_{l}\right|^{2}}\right)_{\widetilde{p}_{l}}^{\mathcal{P}} & l=\hat{l}\\
0 & \mbox{otherwise}
\end{cases}\label{eq:app-tdm-power}
\end{equation}
where the projection yields $\widetilde{p}_{\hat{l}}=\max\{0,\min\{q_{l}/V-1/\left|h_{l}\right|^{2},2^{L\lambda_{\max}}/h_{0}^{2}\}\}$.

3) \emph{Rate allocation: }the rate $\bm{\mu}$ is allocated according
to 
\begin{equation}
\widetilde{\mu}_{l}=\begin{cases}
\log\left(1+\left(\frac{q_{l}\left|h_{l}\right|^{2}}{V}-1\right)_{\widetilde{p}_{l}}^{\mathcal{P}}\right) & l=\hat{l}\\
0 & \mbox{otherwise}
\end{cases}\label{eq:app-tdm-rate}
\end{equation}
Note that the above policy is the solution of the following optimization
problem, 
\begin{eqnarray}
\mbox{maximize} & \qquad\sum q_{l}\log(1+\left|h_{l}\right|^{2}p_{l})-V\sum p_{l}\label{eq:app-tdm-prob}\\
\mbox{subject to} & \mbox{only one link is activated.}\nonumber 
\end{eqnarray}
As a result, the optimum queue-weighted sum transmission rate for
the time division policy is 
\begin{eqnarray}
\sum_{l=1}^{L}q_{l}\widetilde{\mu}_{l} & = & q_{\hat{l}}\log\left(1+\left(\frac{q_{\hat{l}}\left|h_{\hat{l}}\right|^{2}}{V}-1\right)_{\widetilde{p}_{\hat{l}}}^{\mathcal{P}}\right)\geq q_{m}\log\left(1+\left(\frac{q_{m}\left|h_{m}\right|^{2}}{V}-1\right)_{p_{m}}^{\mathcal{P}}\right)\nonumber \\
 & = & \begin{cases}
\begin{array}{c}
0\\
q_{m}\min\left\{ \log\left(\frac{\left|h_{m}\right|^{2}}{V}q_{m}\right),L\lambda_{\max}+\log\frac{|h_{m}|^{2}}{|h_{0}|^{2}}\right\} 
\end{array} & \begin{array}{c}
\|\mathbf{q}\|\left|h_{m}\right|^{2}\leq V\\
\|\mathbf{q}\|\left|h_{m}\right|^{2}>V
\end{array}\end{cases}\nonumber \\
 & \geq & \begin{cases}
\begin{array}{c}
0\\
\|\mathbf{q}\|\min\left\{ \log\left(\frac{S_{\min}}{V}\|\mathbf{q}\|\right),L\lambda_{\max}+\log\frac{S_{\min}}{|h_{0}|^{2}}\right\} 
\end{array} & \begin{array}{c}
\|\mathbf{q}\|S_{\min}\leq V\\
\|\mathbf{q}\|S_{\min}>V
\end{array}\end{cases}\label{eq:app-optimal-power-TDMA}
\end{eqnarray}
where $q_{m}=\|\mathbf{q}\|$ stands for the queue that has the largest
backlog (i.e., $m=\arg\max_{l}\left\{ q_{l}\right\} $). The optimal
utility for the time division policy is then given by (\ref{eq:app-optimal-power-TDMA})
for $\|\mathbf{q}\|S_{\min}>V$ where $P_{t}=\sum\widetilde{p}_{l}=p_{\hat{l}}$
is the total power.

Since, with the same objective, the optimization domain of the time
division MWQ problem (\ref{eq:app-tdm-prob}) is just a subset of
that of the original MWQ problem in (\ref{eq:prob-weighted-queue}),
the MWQ problem (\ref{eq:prob-weighted-queue}) yields a utility $U^{*}=\sum q_{l}\mu_{l}^{*}-V\sum p_{l}^{*}\geq\widetilde{U}$.
To evaluate the queue-weighted utility $\sum q_{l}\mu_{l}^{*}$, we
consider the following two cases.

\emph{Case 1:} When $\sum p_{l}^{*}\geq P_{t}=\sum\widetilde{p}_{l}$,
it is obvious that, for $\|\mathbf{q}\|S_{\min}>V$, 
\[
\sum q_{l}\mu_{l}^{*}\geq\sum q_{l}\widetilde{\mu}_{l}\geq\|\mathbf{q}\|\min\left\{ \log\left(\frac{S_{\min}}{V}\|\mathbf{q}\|\right),L\lambda_{\max}+\log\frac{S_{\min}}{|h_{0}|^{2}}\right\} .
\]

\emph{Case 2:} When $\sum p_{l}^{*}<P_{t}$, we let $V^{'}=\frac{\sum p_{l}^{*}}{P_{t}}V<V$.
Note that decreasing the tradeoff parameter $V$ will increase the
power allocation and hence increase the queue-weighted utility $\sum q_{l}\mu_{l}^{*}$.
Specifically, the optimal utility becomes 
\[
U^{*}=\sum q_{l}\mu_{l}^{*}-V\sum p_{l}^{*}=\sum q_{l}\mu_{l}^{*}-V^{'}P_{t}\geq\sum q_{l}\widetilde{\mu}_{l}^{'}-V^{'}P_{t}\geq\sum q_{l}\widetilde{\mu}_{l}^{'}-V^{'}\sum\widetilde{p}_{l}^{'}
\]
as 
\[
\widetilde{p}_{l}^{'}=\begin{cases}
\left(\frac{q_{l}}{V^{'}}-\frac{1}{\left|h_{l}\right|^{2}}\right)_{\widetilde{p}_{l}^{'}}^{\mathcal{P}}\geq\widetilde{p}_{l}=P_{t} & l=\hat{l}\\
0 & \mbox{otherwise}
\end{cases}
\]
where $\widetilde{\mu}_{l}^{'}$ and $\widetilde{p}_{l}^{'}$ are
the solutions to the time division MWQ problem (\ref{eq:app-tdm-prob}).
Hence $\sum q_{l}\mu_{l}^{*}\geq\sum q_{l}\widetilde{\mu}_{l}^{'}\geq\|\mathbf{q}\|\log\left(\frac{S_{\min}}{V^{'}}\|\mathbf{q}\|\right)\geq\|\mathbf{q}\|\min\left\{ \log\left(\frac{S_{\min}}{V}\|\mathbf{q}\|\right),L\lambda_{\max}+\log\frac{S_{\min}}{|h_{0}|^{2}}\right\} $,
for $\|\mathbf{q}\|S_{\min}>V$.

Combining the above two cases, we prove the inequality (\ref{eq:lem_q-mu-sol-property}).

In addition, as $L\|\mathbf{q}\|\|\bm{\mu}^{*}\|\geq\sum_{l=1}^{L}q_{l}\mu_{l}^{*}\geq\|\mathbf{q}\|\min\left\{ \log\left(\frac{S_{\min}}{V}\|\mathbf{q}\|\right),L\lambda_{\max}+\log\frac{S_{\min}}{|h_{0}|^{2}}\right\} $,
we have $\|\bm{\mu}^{*}\|\geq\frac{1}{L}\min\left\{ \log\left(\frac{S_{\min}}{V}\|\mathbf{q}\|\right),L\lambda_{\max}+\log\frac{S_{\min}}{|h_{0}|^{2}}\right\} $,
for $\|\mathbf{q}\|S_{\min}>V$. Similarly, we can get $\|\bm{\mu}^{*}\|\leq\log\left(\frac{S_{\max}}{V}\|\mathbf{q}\|\right)$.
Hence we prove inequality (\ref{eq:lem_mu-property}).
\end{proof}

\section{Proof of Lemma \ref{lem:error-trans-rate} \label{app:prof-lem-trans-err}}
\begin{proof}
According to (\ref{eq:mu-hat-k}) in Lemma \ref{lem:Optimal-rate-allocation},
$\hat{\mu}_{\pi(k)}=\log\left(\rho_{k}(\mathbf{p})\right)$ under
some permutation $\bm{\pi}$, where 
\[
\rho_{k}(\mathbf{p})=\frac{1+\sum_{i=1}^{k}\left|h_{\pi(i)}\right|^{2}p_{\pi(i)}}{1+\sum_{i=1}^{k-1}\left|h_{\pi(i)}\right|^{2}p_{\pi(i)}}.
\]
Notice that $\rho_{k}(\mathbf{p})$ is a ratio of two polynomials.
In addition, the coefficients $\left|h_{\pi(i)}\right|^{2}$ are bounded
by $S_{\min}$ and $S_{\max}$. Hence $\rho_{k}(\mathbf{p})$ is Lipschitz
continuous, i.e., there exists $0<\beta_{k}<\infty$ depending on
$S_{\min}$ and $S_{\max}$ such that 
\[
\|\rho_{k}(\mathbf{p})-\rho(\mathbf{p}^{*})\|\leq\beta_{k}\|\mathbf{p}-\mathbf{p}^{*}\|=\beta_{k}\|\mathbf{p}_{e}\|.
\]
Therefore, as $\rho_{k}(\centerdot)\geq1$, assuming $\rho_{k}(\mathbf{p})\geq\rho(\mathbf{p}^{*})$,
we have 
\begin{eqnarray*}
\|\hat{\mu}_{\pi(k)}(\mathbf{p})-\hat{\mu}_{\pi(k)}(\mathbf{p}^{*})\| & = & \log\left(\rho_{k}(\mathbf{p})\right)-\log\left(\rho_{k}(\mathbf{p}^{*})\right)\\
 & = & \log\left(1+\frac{\rho_{k}(\mathbf{p})-\rho_{k}(\mathbf{p}^{*})}{\rho_{k}(\mathbf{p}^{*})}\right)\leq\log\left(1+\frac{\beta_{k}\|\mathbf{p}_{e}\|}{\rho_{k}(\mathbf{p}^{*})}\right)\leq\log\left(1+\beta_{k}\|\mathbf{p}_{e}\|\right).
\end{eqnarray*}
Similarly, when $\rho_{k}(\mathbf{p})<\rho(\mathbf{p}^{*})$, we have
\begin{eqnarray*}
\|\hat{\mu}_{\pi(k)}(\mathbf{p})-\hat{\mu}_{\pi(k)}(\mathbf{p}^{*})\| & = & \log\left(\rho_{k}(\mathbf{p}^{*})\right)-\log\left(\rho_{k}(\mathbf{p})\right)\\
 & = & \log\left(1+\frac{\rho_{k}(\mathbf{p}^{*})-\rho_{k}(\mathbf{p})}{\rho_{k}(\mathbf{p})}\right)\leq\log\left(1+\frac{\beta_{k}\|\mathbf{p}_{e}\|}{\rho_{k}(\mathbf{p})}\right)\leq\log\left(1+\beta_{k}\|\mathbf{p}_{e}\|\right).
\end{eqnarray*}
Hence $\|\hat{\bm{\mu}}(\mathbf{p})-\hat{\bm{\mu}}(\mathbf{p}^{*})\|\leq\log\left(1+\beta\|\mathbf{p}_{e}\|\right)$,
where $\beta=\max_{k=\{1,\dots,L\}}\beta_{k}$. Using the triangular
inequality, we obtain 
\[
\|\hat{\bm{\mu}}(\mathbf{p}^{*})\|-\log\left(1+\beta\|\mathbf{p}_{e}\|\right)\leq\|\bm{\hat{\mu}}(\mathbf{p})\|\leq\|\hat{\bm{\mu}}(\mathbf{p}^{*})\|+\log\left(1+\beta\|\mathbf{p}_{e}\|\right)
\]
that leads to the result.
\end{proof}

\section{Proof of Lemma \ref{lem:Lyapunov-drift-property}\label{app:prof-lem-Lyapunov-drift-property}}
\begin{proof}
From Lemma \ref{lem:Optimal-rate-allocation} and \ref{lem:error-trans-rate},
we have 
\begin{eqnarray*}
\sum q_{l}\hat{\mu}_{l}(t) & \geq & \sum q_{l}\left[\mu_{l}^{*}-\log\left(1+\beta\|\mathbf{p}_{e}\|\right)\right]^{+}\\
 & \geq & \sum q_{l}\mu_{l}^{*}-\sum q_{l}\log\left(1+\beta\|\mathbf{p}_{e}\|\right)\\
 & \geq & \|\mathbf{q}\|\min\left\{ \log\left(\frac{S_{\min}}{V}\|\mathbf{q}\|\right),L\lambda_{\max}+\log\frac{S_{\min}}{|h_{0}|^{2}}\right\} -L\|\mathbf{q}\|\log\left(1+\beta\|\mathbf{p}_{e}\|\right)
\end{eqnarray*}

From the optimality condition \cite{Boyd:2004kx} for a convex problem,
we also have $\mathbf{p}_{e}^{T}f(\mathbf{p}_{e};\mathbf{h},\mathbf{q})\leq0$
for all $\mathbf{p}_{e}$. According to the proof of Lemma \ref{lem:Solution-property-MWQ}
in Appendix \ref{app:prof-lem-sol-property}, two cases for $\mathbf{q}(t)$
should be considered. 

\emph{Case 1}: $\|\mathbf{q}\|S_{\min}>V$. The  stochastic Lyapunov
drift (\ref{eq:LV-VSDS}) can be written as 
\begin{eqnarray}
LV(\mathbf{z}) & \leq & -2\kappa\alpha\|\mathbf{p}_{e}\|^{2}+2\gamma_{q}\|\mathbf{p}_{e}\|\log\left(\frac{S_{\max}}{V}\|\mathbf{q}\|\right)+2\gamma_{q}\|\mathbf{p}_{e}\|\log\left(1+\beta\|\mathbf{p}_{e}\|\right)\nonumber \\
 &  & \qquad+\gamma_{h}a_{A}\|\mathbf{p}_{e}\|\|\mathbf{h}\|-a_{A}\|\mathbf{h}\|^{2}+2\gamma_{q}\lambda_{\max}\|\mathbf{p}_{e}\|+2L\|\mathbf{q}\|\log\left(1+\beta\|\mathbf{p}_{e}\|\right)\nonumber \\
 &  & \qquad-2\|\mathbf{q}\|\min\left\{ \log\left(\frac{S_{\min}}{V}\|\mathbf{q}\|\right),L\lambda_{\max}+\log\frac{S_{\min}}{|h_{0}|^{2}}\right\} +2L\|\mathbf{q}\|\lambda_{\max}+C\label{eq:LV-3}
\end{eqnarray}
where 
\begin{eqnarray*}
\mbox{tr}\left(2\left(A^{\frac{1}{2}}\right)^{T}\psi_{h}^{T}\psi_{h}A^{\frac{1}{2}}+\psi_{q}^{T}\psi_{q}\right)+\sum_{l=1}^{L}\left(2a_{l}+\lambda_{l}\right) & \leq & \sum_{l=1}^{L}2a_{l}\left(1+\gamma_{h}^{2}\right)+L\gamma_{q}^{2}+\sum_{l=1}^{L}\lambda_{l}\\
 & \leq & L(2a_{A}(1+\gamma_{h}^{2})+\gamma_{q}^{2}+\lambda_{\max})\\
 & \triangleq & C
\end{eqnarray*}

To find the upper bound of the R.H.S. of (\ref{eq:LV-3}), we divide
it into 2 parts as follows.
\begin{eqnarray*}
I_{1} & = & -\kappa\alpha\|\mathbf{p}_{e}\|^{2}+2\gamma_{q}\lambda_{\max}\|\mathbf{p}_{e}\|-\|\mathbf{q}\|\min\left\{ \log\left(\frac{S_{\min}}{V}\|\mathbf{q}\|\right),L\lambda_{\max}+\log\frac{S_{\min}}{|h_{0}|^{2}}\right\} \\
 &  & \qquad+L\lambda_{\max}\|\mathbf{q}\|+2\gamma_{h}a_{A}\|\mathbf{p}_{e}\|\|\mathbf{h}\|-2a_{A}\|\mathbf{h}\|^{2}+C,\\
I_{2} & = & -\kappa\alpha\|\mathbf{p}_{e}\|^{2}+2\gamma_{q}\|\mathbf{p}_{e}\|\log\left(\frac{S_{\max}}{V}\|\mathbf{q}\|\right)+2\gamma_{q}\|\mathbf{p}_{e}\|\log\left(1+\beta\|\mathbf{p}_{e}\|\right)+L\lambda_{\max}\|\mathbf{q}\|\\
 &  & \qquad+2L\|\mathbf{q}\|\log\left(1+\beta\|\mathbf{p}_{e}\|\right)-\|\mathbf{q}\|\min\left\{ \log\left(\frac{S_{\min}}{V}\|\mathbf{q}\|\right),L\lambda_{\max}+\log\frac{S_{\min}}{|h_{0}|^{2}}\right\} .
\end{eqnarray*}

(1) With some calculations, it is not difficult to show that $I_{1}\leq-\|\mathbf{h}\|+\frac{\gamma_{q}^{2}\lambda_{\max}^{2}}{\kappa\alpha}+\frac{V}{S_{\min}}2^{L\lambda_{\max}-1}+\frac{1}{8a_{A}[1-\gamma_{h}^{2}a_{A}/(k\alpha)]}+C$,
for $\kappa>\frac{2}{\alpha}\max\left\{ \gamma_{q}^{2},\beta\gamma_{q}\right\} $.

(2) Denote $g_{1}(S_{\min},S_{\max})=\max_{\{\|\mathbf{p}_{e}\|,\|\mathbf{q}\|\}}\left\{ I_{2}+\left(\|\mathbf{p}_{e}\|+\|\mathbf{q}\|\right)\right\} $.
We can easily find that $g_{1}$ is bounded above for all $S_{\min}$
and $S_{\max}$ in the domain%
\footnote{To show a real valued function $f(x,y)$ is bounded above, we start
from a point ($x_{0},y_{0})$ in the domain and proceed to show that,
by substituting with $y=x_{0}+\beta(y-y_{0})$, $f(x,y(x;\beta))$
is bounded above uniformly for every $\beta\in\mathbb{R}$. It can
be verified that $f(x,y(x;\beta))$ satisfies this condition in our
case.%
}. Note that an upper bound expression for $g_{1}$ is always obtainable,
since it is only a simple bivariate programming problem. Therefore,
we obtain $I_{2}\leq-\left(\|\mathbf{p}_{e}\|+\|\mathbf{q}\|\right)+g_{1}(S_{\min},S_{\max})$. 

As a result, we have 
\begin{eqnarray*}
LV & \leq & -\left(\|\mathbf{p}_{e}\|+\|\mathbf{q}\|+\|\mathbf{h}\|\right)+\frac{\gamma_{q}^{2}\lambda_{\max}^{2}}{\kappa\alpha}+\frac{V}{S_{\min}}2^{L\lambda_{\max}-1}\\
 &  & \qquad+\frac{1}{8a_{A}[1-\gamma_{h}^{2}a_{A}/(k\alpha)]}+g_{1}(S_{\min},S_{\max})+C.
\end{eqnarray*}

\emph{Case }2: $\|\mathbf{q}\|S_{\min}\leq V$. Here we have $\|\mathbf{q}\|\leq\frac{V}{S_{\min}}$.
From the property in Appendix \ref{app:prof-lem-sol-property}, the
stochastic Lyapunov drift (\ref{eq:LV-VSDS}) can be written as
\begin{eqnarray*}
LV(\mathbf{z}) & \leq & -2\kappa\alpha\|\mathbf{p}_{e}\|^{2}+2\gamma_{q}\|\mathbf{p}_{e}\|\log\left(S_{\max}S_{\min}\right)+2\gamma_{q}\|\mathbf{p}_{e}\|\log\left(1+\beta\|\mathbf{p}_{e}\|\right)+\gamma_{h}a_{A}\|\mathbf{p}_{e}\|\|\mathbf{h}\|\\
 &  & \qquad+2\gamma_{q}\lambda_{\max}\|\mathbf{p}_{e}\|-a_{A}\|\mathbf{h}\|^{2}+\frac{V}{S_{\min}}\log\left(1+\beta\|\mathbf{p}_{e}\|\right)+\frac{V}{S_{\min}}L\lambda_{\max}+C\\
 & \leq & -\left(\|\mathbf{p}_{e}\|+\|\mathbf{h}\|\right)+g_{2}(S_{\min},S_{\max})+\frac{\gamma_{q}^{2}\lambda_{\max}^{2}}{\kappa\alpha}\\
 &  & \qquad+\frac{V}{S_{\min}}L\lambda_{\max}+\frac{1}{8a_{A}[1-\gamma_{h}^{2}a_{A}/(k\alpha)]}+C\\
 & \leq & -\left(\|\mathbf{p}_{e}\|+\|\mathbf{h}\|\right)+J_{0}+\frac{\gamma_{q}^{2}\lambda_{\max}^{2}}{\kappa\alpha}+\frac{V}{S_{\min}}L\lambda_{\max}\\
 &  & \qquad+\frac{1}{8a_{A}[1-\gamma_{h}^{2}a_{A}/(k\alpha)]}+C-\|\mathbf{q}\|+\frac{V}{S_{\min}}
\end{eqnarray*}
where 
\begin{eqnarray*}
g_{2}(S_{\min},S_{\max}) & = & \max\{-\kappa\alpha\|\mathbf{p}_{e}\|^{2}+2\gamma_{q}\|\mathbf{p}_{e}\|\log\left(S_{\max}S_{\min}\right)\\
 &  & \qquad\qquad+2\gamma_{q}\|\mathbf{p}_{e}\|\log\left(1+\beta\|\mathbf{p}_{e}\|\right)+\frac{V}{S_{\min}}\log\left(1+\beta\|\mathbf{p}_{e}\|\right)\}.
\end{eqnarray*}
Therefore, we have {[}since $C=L(2a_{A}(1+\gamma_{h}^{2})+\gamma_{q}^{2}+\lambda_{\max})${]}
\begin{eqnarray*}
LV(\mathbf{z}) & \leq & -\left(\|\mathbf{p}_{e}\|+\|\mathbf{q}\|+\|\mathbf{\mathbf{h}}\|\right)+L(2a_{A}(1+\gamma_{h}^{2})+\gamma_{q}^{2}+\lambda_{\max})\\
 &  & \qquad+\frac{\gamma_{q}^{2}\lambda_{\max}^{2}}{\kappa\alpha}+\frac{V}{S_{\min}}2^{L\lambda_{\max}-1}+\frac{1}{8a_{A}[1-\gamma_{h}^{2}a_{A}/(k\alpha)]}+g(S_{\min},S_{\max})
\end{eqnarray*}
where $g(S_{\min},S_{\max})=\max\left\{ g_{1}(S_{\min},S_{\max}),g_{1}(S_{\min},S_{\max})\right\} $. 
\end{proof}

\section{Proof of Theorem \ref{thm:convergence-compensation} \label{app:Proof-thm-convergence-compensation}}
\begin{proof}
Consider the virtual error dynamic system 
\begin{eqnarray*}
d\mathbf{p}_{e} & = & \kappa\nabla_{\mathbf{p}}\mathcal{L}\left(\hat{\bm{\mu}}(\mathbf{p}),\mathbf{p};\mathbf{h},\mathbf{q}\right)dt+\left(\hat{\varphi}_{q}(\mathbf{p};\centerdot)-\hat{\varphi}_{q}(\mathbf{p}^{*};\centerdot)\right)d\mathbf{q}+\mbox{Re}\left(\hat{\varphi}_{h}(\mathbf{p};\centerdot)d\mathbf{h}-\hat{\varphi}_{h}(\mathbf{p}^{*};\centerdot)d\mathbf{h}\right)\\
 & = & \left[\kappa\nabla_{\mathbf{p}}\mathcal{L}\left(\hat{\bm{\mu}}(\mathbf{p}),\mathbf{p};\mathbf{h},\mathbf{q}\right)-\bm{\mu}(t)\left(\hat{\varphi}_{q}(\mathbf{p};\centerdot)-\hat{\varphi}_{q}(\mathbf{p}^{*};\centerdot)\right)-\frac{1}{2}\mbox{Re}\left[\left(\hat{\varphi}_{h}(\mathbf{p};\centerdot)-\hat{\varphi}_{h}(\mathbf{p}^{*};\centerdot)\right)A\mathbf{h}\right]\right]dt\\
 &  & \qquad+\left(\hat{\varphi}_{q}(\mathbf{p};\centerdot)-\hat{\varphi}_{q}(\mathbf{p}^{*};\centerdot)\right)d\mathbf{N}(t)+\mbox{Re}\left[\left(\hat{\varphi}_{h}(\mathbf{p};\centerdot)-\hat{\varphi}_{h}(\mathbf{p}^{*};\centerdot)\right)A^{\frac{1}{2}}d\mathbf{W}(t)\right].
\end{eqnarray*}
Taking the Lyapunov function as $V(\mathbf{p}_{e})=\frac{1}{2}\mathbf{p}_{e}^{T}\mathbf{p}_{e}$.
The Lyapunov drift is defined as $LV(\mathbf{p}_{e})=\lim_{\delta\downarrow0}\frac{1}{\delta}\left\{ \mathbb{E}\left[V\left(\mathbf{p}_{e}(t+\delta)\right)|\mathbf{p}_{e}(t)\right]-V\left(\mathbf{p}_{e}(t)\right)\right\} $.
Note that $\mathbb{E}\left[h_{l}\right]=0$. The drift can be derived
into 
\begin{eqnarray*}
LV(\mathbf{p}_{e}) & = & \mathbf{p}_{e}^{T}f(\mathbf{p}_{e};\centerdot)-\mathbf{p}_{e}^{T}\bm{\mu}(t)\left(\hat{\varphi}_{q}(\mathbf{p};\centerdot)-\hat{\varphi}_{q}(\mathbf{p}^{*};\centerdot)\right)+\sum_{l=1}^{L}\lambda_{l}\mathbf{p}_{e}^{T}\left(\hat{\varphi}_{q}^{(l)}(\mathbf{p};\centerdot)-\hat{\varphi}_{q}^{(l)}(\mathbf{p}^{*};\centerdot)\right)\\
 &  & \;+\frac{1}{2}\mbox{tr}\left[\left(\hat{\varphi}_{q}(\mathbf{p};\centerdot)-\hat{\varphi}_{q}(\mathbf{p}^{*};\centerdot)\right)^{T}\left(\hat{\varphi}_{q}(\mathbf{p};\centerdot)-\hat{\varphi}_{q}(\mathbf{p}^{*};\centerdot)\right)\right]\\
 &  & \;+\frac{1}{2}\mbox{tr}\left[\left(A^{\frac{1}{2}}\right)^{T}\left(\hat{\varphi}_{h}(\mathbf{p};\centerdot)-\hat{\varphi}_{h}(\mathbf{p}^{*};\centerdot)\right)^{T}\left(\hat{\varphi}_{h}(\mathbf{p};\centerdot)-\hat{\varphi}_{h}(\mathbf{p}^{*};\centerdot)\right)A^{\frac{1}{2}}\right]\\
 & \leq & -\kappa\alpha\|\mathbf{p}_{e}\|^{2}+\mu_{\max}L_{q}\|\mathbf{p}_{e}\|^{2}+\lambda_{\max}LL_{q}\|\mathbf{p}_{e}\|^{2}+\frac{1}{2}L_{q}^{2}\|\mathbf{p}_{e}\|^{2}+\frac{1}{2}a_{A}L_{h}^{2}\|\mathbf{p}_{e}\|^{2}\\
 & = & -\rho\|\mathbf{p}_{e}\|^{2}
\end{eqnarray*}
where $\rho=-\kappa\alpha+\mu_{\max}L_{q}+\lambda_{\max}LL_{q}+\frac{1}{2}L_{q}^{2}+\frac{1}{2}a_{A}L_{h}^{2}>0$.
Hence from the asymptotic stochastic stability results given in \cite{Mao:SDE1997}
we have proven the theorem.
\end{proof}

\bibliographystyle{IEEEtran}
\bibliography{Bib_Queue_Resource_Alloc,Bib_Converge_Analysis,/Users/Allen/Dropbox/Draft/Bib_books,/Users/Allen/Dropbox/Draft/Bib_Classical_paper}

\end{document}